\documentclass[floatfix,aps,showpacs,superscriptaddress,amsmath,amssymb,longbibliography]{revtex4-2}

\usepackage[autostyle=true]{csquotes}

\usepackage{times}

\usepackage{mathtools}
\usepackage{textcomp}
\usepackage{gensymb}
\usepackage{graphicx}
\usepackage{siunitx}
\usepackage{ulem}
\usepackage{physics}
\usepackage{appendix}

\usepackage[unicode=true,
 bookmarks=true,bookmarksnumbered=true,bookmarksopen=true,bookmarksopenlevel=2,
 breaklinks=false,pdfborder={0 0 1},backref=false,colorlinks=true]
 {hyperref}

 \hypersetup{linkcolor=blue, citecolor=blue, urlcolor=blue, filecolor=blue, pdfpagelayout=OneColumn,
  pdfnewwindow=true, pdfstartview=XYZ, plainpages=false}

\usepackage[all]{hypcap}
\usepackage{color}

\makeatletter
\def\mathcolor#1#{\@mathcolor{#1}}%
\def\@mathcolor#1#2#3{%
  \protect\leavevmode%
  \begingroup\color#1{#2}#3\endgroup%
}%
\newcommand{\msout}[1]{\text{\sout{\ensuremath{#1}}}}%

\newcommand{\modif}{}
\newcommand{\highlight}{}

\newcommand{\sam}[2]{%
\ifmmode%
  \msout{#1}\mathcolor{red}{#2}%
\else%
  \sout{#1}\textcolor{red}{#2}%
\fi}
\newcommand{\samO}[2]{%
\ifmmode%
  \msout{#1}\mathcolor{magenta}{#2}%
\else%
  \sout{#1}\textcolor{magenta}{#2}%
\fi}

\newcommand{\cyr}[2]{%
\ifmmode%
  \msout{#1}\mathcolor{red}{#2}%
\else%
  \sout{#1}\textcolor{red}{#2}%
\fi}

\newcommand{\gab}[2]{%
\ifmmode%
  \msout{#1}\mathcolor{green}{#2}%
\else%
  \sout{#1}\textcolor{green}{#2}%
\fi}

\makeatother

\makeatletter

\newcommand\Autoref[1]{\@first@ref#1,@}
\def\@throw@dot#1.#2@{#1}
\def\@set@refname#1{
    \edef\@tmp{\getrefbykeydefault{#1}{anchor}{}}%
    \xdef\@tmp{\expandafter\@throw@dot\@tmp.@}%
    \ltx@IfUndefined{\@tmp autorefnameplural}%
         {\def\@refname{\@nameuse{\@tmp autorefname}s}}%
         {\def\@refname{\@nameuse{\@tmp autorefnameplural}}}%
}
\def\@first@ref#1,#2{%
  \ifx#2@\autoref{#1}\let\@nextref\@gobble
  \else%
    \@set@refname{#1}
    \@refname~\ref{#1}
    \let\@nextref\@next@ref
  \fi%
  \@nextref#2%
}
\def\@next@ref#1,#2{%
   \ifx#2@ and~\ref{#1}\let\@nextref\@gobble
   \else, \ref{#1}
   \fi%
   \@nextref#2%
}

\makeatother

\begin{document}

\def\be{\begin{equation}}
\def\ee{\end{equation}}

\def\myvec#1{{\bf #1}}
\def\Esw{\myvec{E}_{sw}}
\def\Hsw{\myvec{H}_{sw}}

\title{Harvesting information to control non-equilibrium states of active matter}

\author{R\'emi Goerlich}
\affiliation{ Universit\'e de Strasbourg, CNRS, Institut de Physique et Chimie des Mat\'eriaux de Strasbourg, UMR 7504, F-67000 Strasbourg, France}
\affiliation{Universit\'e de Strasbourg, CNRS, Institut de Science et d'Ing\'enierie Supramol\'eculaires, UMR 7006, F-67000 Strasbourg, France}

\author{Lu\'{\i}s Barbosa Pires}
\affiliation{Universit\'e de Strasbourg, CNRS, Institut de Science et d'Ing\'enierie Supramol\'eculaires, UMR 7006, F-67000 Strasbourg, France}

\author{Giovanni Manfredi}
\email{giovanni.manfredi@ipcms.unistra.fr}
\affiliation{ Universit\'e de Strasbourg, CNRS, Institut de Physique et Chimie des Mat\'eriaux de Strasbourg, UMR 7504, F-67000 Strasbourg, France}

\author{Paul-Antoine Hervieux}
\affiliation{ Universit\'e de Strasbourg, CNRS, Institut de Physique et Chimie des Mat\'eriaux de Strasbourg, UMR 7504, F-67000 Strasbourg, France}

\author{Cyriaque Genet}
\email{genet@unistra.fr}
\affiliation{Universit\'e de Strasbourg, CNRS, Institut de Science et d'Ing\'enierie Supramol\'eculaires, UMR 7006, F-67000 Strasbourg, France}

\date{\today}

\begin{abstract}
We propose to use a correlated noise bath to drive an optically trapped Brownian particle that mimics active biological matter. Thanks to the flexibility and precision of our setup, we are able to control the different parameters that drive the stochastic motion of the particle with unprecedented accuracy, thus reaching strongly correlated regimes that are not easily accessible with real active matter.
In particular, by using the correlation time (i.e., the ``color") of the noise as a control parameter, we can trigger transitions between two non-equilibrium steady states with no expended work, but only a calorific cost. Remarkably, the measured heat production is directly proportional to the spectral entropy of the correlated noise, in a fashion that is reminiscent of Landauer's principle. Our procedure can be viewed as a method for harvesting information from the active fluctuations.
\end{abstract}

\maketitle

\section{Introduction}
Chemical and biological non-equilibrium processes reveal the special role played by fluctuations at mesoscopic scales, raising fascinating questions that form a major topic of current transdisciplinary research \cite{Rao2016,Fang2019,Davis2020}. The combined development of stochastic thermodynamics and optical trapping experiments offers an appropriate framework to describe such processes \cite{Bustamante2005,CilibertoPRX2017}. Recently, active fluctuations were injected inside optical traps by adding correlated noises to the trapping potential \cite{Park2020,Paneru2021,Militaru2021}, thus providing optomechanical models of active matter that can be studied in close relation with theory \cite{Marconi2017,Joanny2017,VanWijland2021}.

In this work, we follow this approach to decipher the energetics engaged between non-equilibrium states and correlated baths \cite{Eichhorn2019, Speck2016, Speck2021}. To do so, we bring an optically confined microparticle in an active-matter-like non-equilibrium steady state (NESS). The injected correlated noise yields the constant external source of heat needed to maintain the particle in a NESS. We monitor the non-Brownian diffusion of the trapped particle and probe large deviations from thermal equilibrium in strongly correlated regimes unexplored so far with real, biological active matter \cite{Volpe2016,DiLeonardo2014,Maggi2017}.
\highlight{The platform we propose can be used to explore experimentally the fascinating features of a large variety of systems driven by non-trivial baths that have been recently described theoretically \cite{Loi2008, Underhill2008, Graneck2021, Ye2020, Foffano2012}.
It can also be applied to different tracers \cite{Belan2021} such as molecular motors \cite{Ariga2018, Speck2021}, or to many-body systems \cite{VolpeReview2016, Abbaspour2021}, in order to study their response to arbitrary baths.}

We further demonstrate that it is possible to drive the particle from one NESS to another through the sole change of the correlation time (i.e., the ``color'') of the active fluctuations. For instance, transitions between NESS are known to occur when biological matter undergoes a change in mechanical properties, such as during mitosis, and thereby a modification of the intracellular noise spectrum \cite{Hurst2021, Taubenberger2020}. Modulating the color of the noise without changing its amplitude makes it possible to achieve heat production at constant energy input. We show that this capacity to induce NESS-to-NESS transitions at zero energetic cost is rooted in the informational content of active fluctuations. Indeed, correlated noise carries information that can be quantified by the spectral entropy $H_S$ \cite{Zaccarelli2013}, the counterpart of Shannon's entropy in the frequency domain, which  depends on the noise correlation time. Remarkably, we find that, for a constant noise amplitude quantified by an effective \modif{ equipartition-based temperature $T_{\rm eq}$, the heat $\Delta Q$ generated through the transition is proportional to the entropy of the correlated noise: $\Delta Q=k_{B}T_{\rm eq}\Delta H_S$,  in a fashion that is reminiscent of Landauer's principle \cite{Landauer1961,Frank2018}.}
Effectively, our protocol harvests information from the colored noise and turns it into heat released to the surrounding fluid throughout the NESS-to-NESS transition. Our result reveals, \highlight{in the broad context of exponentially correlated noises,} a deep connection between information and non-equilibrium thermodynamics, which is central to molecular motors and living systems capable of extracting energy from their fluctuating environments to accelerate their average motion \cite{Astumian1998,Ezber2020,Linke2020}.

\begin{figure}[htb!]
	\centering{
		\includegraphics[width=0.8\linewidth]{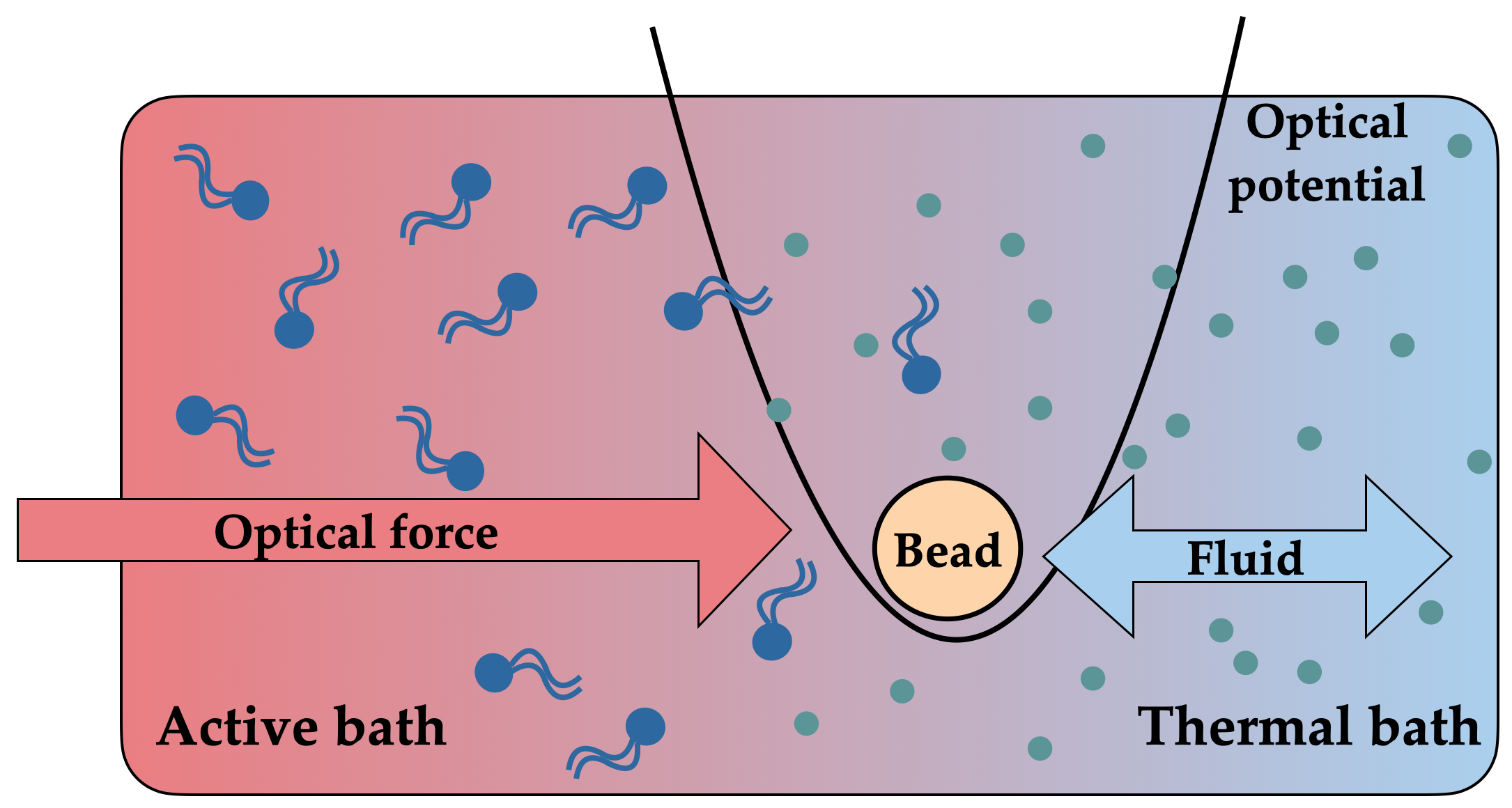}}
	\caption{Schematic view of the experimental realization. The particle is trapped by the optical potential generated by a focused laser beam. Its stochastic motion is driven by two random baths: a thermal (white noise) bath associated with the surrounding fluid at room temperature $T=296$ K (blue)  and an active (colored noise) bath generated by an additional laser exerting on the particle an actively fluctuating radiation pressure force (red).
}
	\label{fig:Analogy}
\end{figure}

\section{Experimental realization}
Our experiment consists in optically trapping a single $3~\si{\micro \meter}$ dielectric bead with a $785~\si{\nano\meter}$  Gaussian laser beam.
The optical potential created by the gradient forces at the waist of the beam is harmonic \highlight{for small displacements}, with a stiffness that is proportional to the intensity of the laser.
The bead is immersed in water at ambient temperature and undergoes random motion due to the thermal fluctuations consistent with an Ornstein-Uhlenbeck process.
An additional radiation pressure force is applied to the bead by a second laser, whose intensity is digitally controlled through time by an acousto-optic modulator.
Any waveform can be sent on the bead, including noise with arbitrary spectrum, up to $\approx 10^5 \rm Hz$.
A schematic view of the experimental realization is shown in Fig. \ref{fig:Analogy} and the experimental setup is described in the Appendix \ref{APPENDIX_samples}.

Active particles are characterized by a persistence that can be mimicked by an exponentially correlated Gaussian noise \cite{Szamel2014, VanWijland2016, VolpeReview2016} generated by an Ornstein-Uhlenbeck process \cite{Hanggi1987, Miguel1980}
\be
d\eta_t = -\omega_c \, \eta_t dt + \sqrt{2 \alpha\omega_c} \, dW_t,
\label{eq:noise}
\ee
where $dW_t$ is a $\delta$-correlated Wiener process. Equation \eqref{eq:noise} yields a noise $\eta_t$ with variance $\alpha$ \footnote{We stress that the amplitude coefficient $\alpha$ for the colored noise $\eta_t$ is fixed digitally: its impact on the motional dynamics described below by Eq. (\ref{eq:model}) is taken over by the active diffusivity $D_a$. However, as is clear from in Eq. (\ref{eq:noise}), it has a fixed dimension (Hz) and thus will appear in the evaluation of the MSD.} and correlation $\langle \eta_t \eta_s \rangle = \alpha e^{-|t -s|\omega_c}$ , where $\omega_c$ is the inverse of the correlation time $\tau_c$.
We emphasize the great flexibility of this model, which enables us to tune independently the amplitude and correlation of the noise.

The correlated noise $\eta_t$ is sent to the bead via the radiation pressure of the secondary laser beam -- see Appendix \ref{APPENDIX_ExtNoise}.
Hence, the overall motion of the bead is subjected to three forces: the optical trapping force, the white noise thermal bath, and the colored noise applied by the radiation pressure.
The position $x_t$ of the bead obeys the Langevin equation \cite{Szamel2014, DiLeonardo2014}
\be
\dot{x}_t = - \omega_0 x_t + \sqrt{2 D}\xi_t + \sqrt{2 D_a}\eta_t ,
\label{eq:MainModel}
\ee
where $\omega_0 = \kappa / \gamma$ is the inverse of the relaxation time in the trap, with $\kappa$ being its stiffness and $\gamma$ the Stokes viscous drag.  $\xi_t$ is a white noise that models the fluid thermal bath, with diffusion coefficient $D = k_B T/\gamma$, where $k_B$ is the Boltzmann constant and $T$ the fluid temperature. $D_a$ is the \textit{active} diffusivity associated with the colored noise $\eta_t$, which incorporates the optomechanical coupling between the bead and the noisy radiation pressure.
Equation \eqref{eq:MainModel} describes a non-Markovian process that does not respect the fluctuation-dissipation theorem, as the correlated fluctuations $\eta_t$ of the active bath are not compensated by the instantaneous dissipation \cite{Maggi2017}, as discussed in the Appendix \ref{APPENDIX:FDT}.

\section{Out-of-equilibrium properties}
One remarkable feature of active matter is that, due to the correlation properties of the bath, it diffuses in non-Brownian fashion \cite{Joanny2017, DiLeonardo2014, Libchaber2000, Grosberg2018}.
The appropriate observable to estimate the departure from Brownian motion is  the mean square displacement (MSD) $\langle \delta x^2 (\Delta)\rangle = \langle \left( x_{t+\Delta} - x_t \right)^2 \rangle$, where $\Delta$ is a lag time.
In the short-time limit, the bead explores the available space inside the trap according to the free diffusive motion $\langle \delta x^2 (\Delta)\rangle \sim\Delta^\beta$, with $\beta = 1$ for normal and $\beta \neq 1$ for anomalous diffusion  \cite{METZLER2000, Metzler2021}.

\begin{figure}[htb!]
	\centering{
		\includegraphics[width=0.4\textwidth]{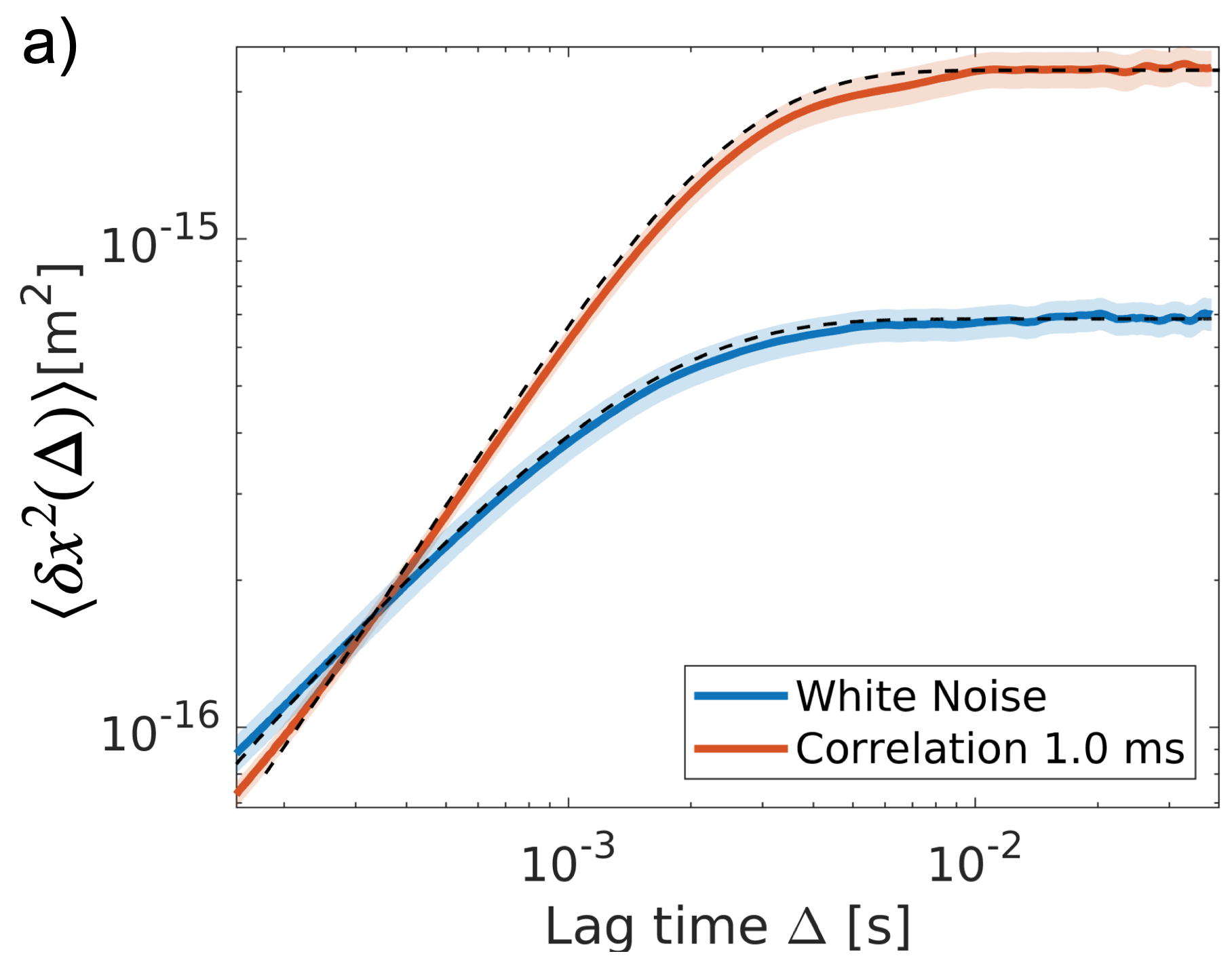}
		\label{fig:MSD}}\\
	\centering{
		\includegraphics[width=0.43\textwidth]{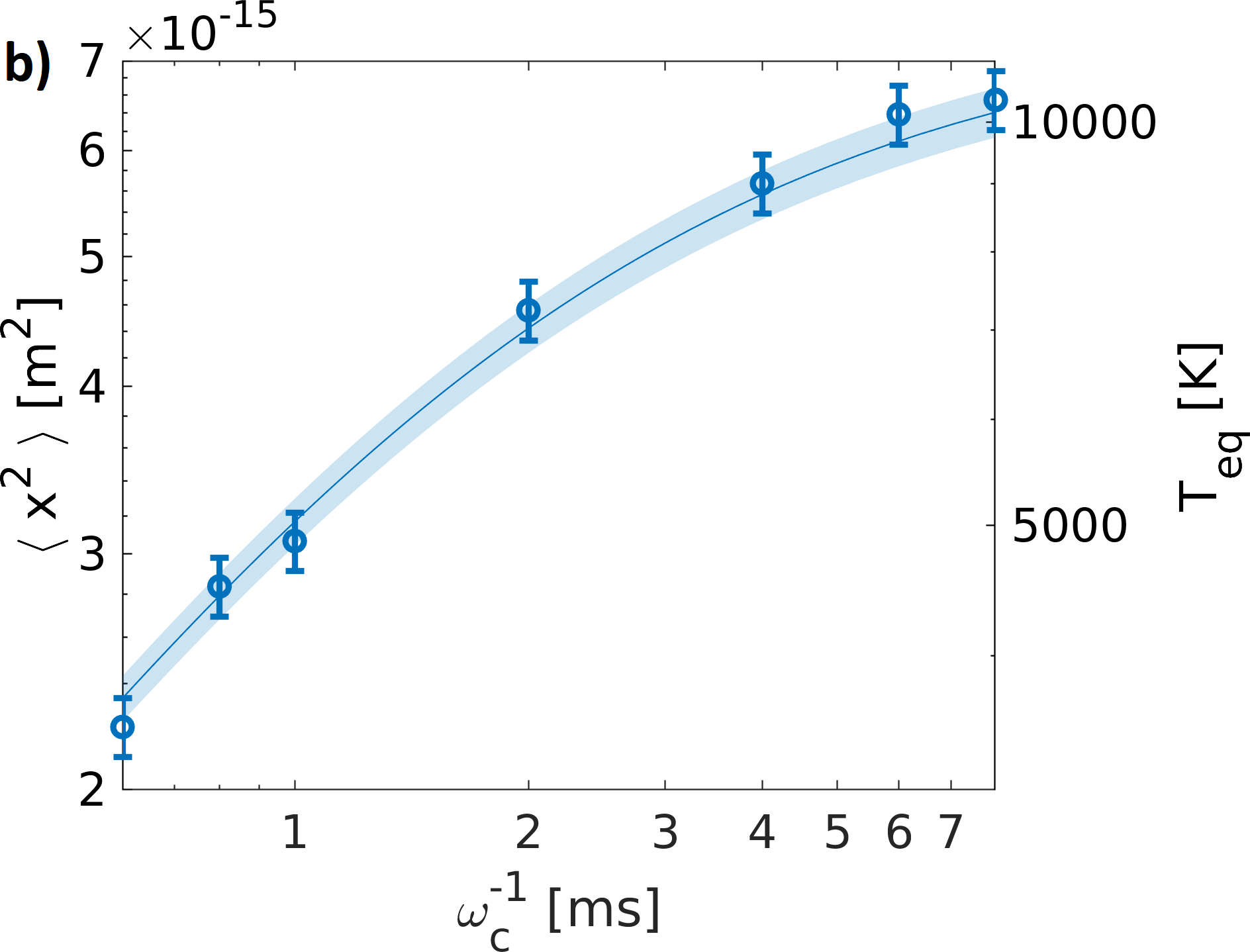}
		\label{fig:Equipart}}
	\caption{ (a) Mean square displacement of the position $x_t$ measured experimentally at room temperature $T=296~\si{\kelvin}$ and relaxation time inside the optical trap $\omega_0^{-1}=1.2~\si{\milli\second}$. The blue line corresponds to a white noise added to the existing thermal fluctuations inside the fluid, showing a short-time diffusive limit ($\sim \Delta^{0.94}$); the red line corresponds to a correlated noise with $\tau_c = 1$ ms, showing superdiffusivity  ($\sim \Delta^{1.5}$) ; \modif{the superimposed dashed lines are fit using analytical MSD, yielding for the white noise case $D_{\textrm{ eq}} = 0.34 > D = 0.17$ $\mu$m$^2$/s (blue line) and for the colored noise case, $D_a = 1.441 \times 10^{4}$ $\mu$m$^2$/s (red line). The shaded regions account for the uncertainties associated with the fitting errors and the systematic error in the sphere radius.}
		(b) Relation between motional variance $\langle x_t^2\rangle$ \modif{(left axis) and correlation time of the bath $\omega_c^{-1}$ ranging from $0.6$ ms to $8$ ms with constant bath noise amplitude. The solid line is the analytical expression of the variance Eq. (\ref{eq:var}), using the average value of $D_a$ extracted from the fit of the MSDs corresponding to each value of $\omega_c^{-1}$. For each case, a temperature $T_{\textrm{eq}}$ (right axis) can be evaluated by multiplying the variance by $\kappa / k_B$ relying on the equipartition theorem. The graph shows clearly that $T_{\textrm{eq}}$ strongly depends on the correlation time of the bath, preventing a simple equiparition relation as explained in the main text. Error bars on the variance and temperature are accounting for both the systematic error on parameters (sphere radius, stiffness, temperature, calibration) and the statistical error on the variance estimation, by a $\chi^2$ test with $3\sigma$ confidence interval. The shaded area accounts, in the analytical evaluation of the variance, for the fitting dispersion of $D_a$.}
}
\end{figure}

In Fig. \ref{fig:MSD}(a), we present the MSD for two processes, obeying Eq. \eqref{eq:model} with the radiation pressure injecting in the trap either a colored noise, or a white noise of equal amplitude.
\modif{In the latter case, the bead diffuses normally and the data can be fitted with the standard Brownian MSD
\be
	\langle \delta x^2 (\Delta)\rangle = \frac{2 D_{\textrm{eq}}}{\omega_0} \left( 1 - e^{-\omega_0 \Delta} \right)
\ee
with an effective diffusion coefficient given by a single temperature $D_{\textrm{eq}} = k_B T_{\textrm{eq}} / \gamma > D$ defined through equipartition, as detailed below.}

In contrast, the trajectory driven by a colored noise is superdiffusive and cannot be \modif{simply described with a standard MSD with a single modified prefactor $D_{\textrm{eq}}$}.
Instead, the experimental MSDs are fitted with a modified expression (see Appendix \ref{APPENDIX_MSD}) taking into account the noise correlation \modif{and involving two independent diffusion coefficients $D$ and $D_a$. The first coefficient $D$ is set by the room temperature $T$, while $D_a$ corresponds to the only fitting parameter.}

\vspace{1mm}

Of course, all regimes with additional noise are characterized by a departure from the thermal equipartition condition $ \langle x_t^2 \rangle = D/\omega_0 = k_B T / \kappa$, as a signature of the non-equilibrium nature of the system \cite{DiLeonardo2014, VanWijland2016}.
\modif{When the  applied noise is white, the equipartition theorem can be retrieved with a temperature defined by the variance $T_{\textrm{eq}} = \kappa \langle x^2 \rangle / k_B$.
This \textit{equipartition temperature} $T_{\textrm{eq}}$ is  a function of the the thermal diffusion coefficient $D$ and the amplitude of the applied noise, only.
In contrast, when the  applied noise is correlated, the variance (see Appendix \ref{APPENDIX_MSD}) is given by
\be
	\langle x^2 \rangle = \frac{D}{\omega_0} + \frac{D_a \alpha}{\omega_0 (\omega_0 + \omega_c)}.
	\label{eq:var}
\ee
Here, the equipartition temperature $T_{\textrm{eq}} = \kappa \langle x^2 \rangle / k_B$ is not only a function of the noise amplitude $D_a$ but also of its correlation time $\omega_c^{-1}$ and $T_{\textrm{eq}}(D_a, \omega_c) \neq \gamma (D + D_a)/k_B $. }\\

\modif{On Fig. \ref{fig:Equipart}(b) we represent both the variance (left axis) and the associated equipartition temperature $T_{\textrm{eq}} = \kappa \langle x^2 \rangle / k_B$ as a function of the correlation time $\omega_c^{-1}$ of the bath, for a constant noise amplitude.
The experimental variances and temperatures are well captured by the analytical result Eq. (\ref{eq:var}), within the systematic and statistical errors.
We observe indeed that equipartition takes a more complex character: the temperature is not uniquely related to the magnitude of the driving noise, but also depends on a second variable, its correlation time.
In Appendix \ref{APPENDIX_ExtNoise} we discuss in more details the relations between the variance, equipartition temperature and correlation time for various sets of parameters.
}

\section{Transition protocols between two NESS}
The correlation time of the noise may further be used as a control parameter in a protocol that brings the system from one NESS to another, without changing either the confining potential or the temperature.
The simplest possible protocol is a step-like sudden change of the correlation time $\tau_c$ (referred to as ``STEP protocol" hereafter), while keeping the noise amplitude constant.
We apply the STEP protocol on the system at a low repetition rate, so that it reaches a steady state in between each change of $\tau_c$.
The ergodic hypothesis, carefully verified (see Appendix \ref{APPENDIX_Ergodicity}), leads us to build an ensemble of  $\approx 1.1 \times 10^{4}$ independent trajectories experiencing the same protocol.
The main quantity of interest here will be the variance of the response of the bead $\langle x_t^2 \rangle$ \cite{Goerlich2021}.

\begin{figure}[htb]
	\centering{
		\includegraphics[width=0.35\textwidth]{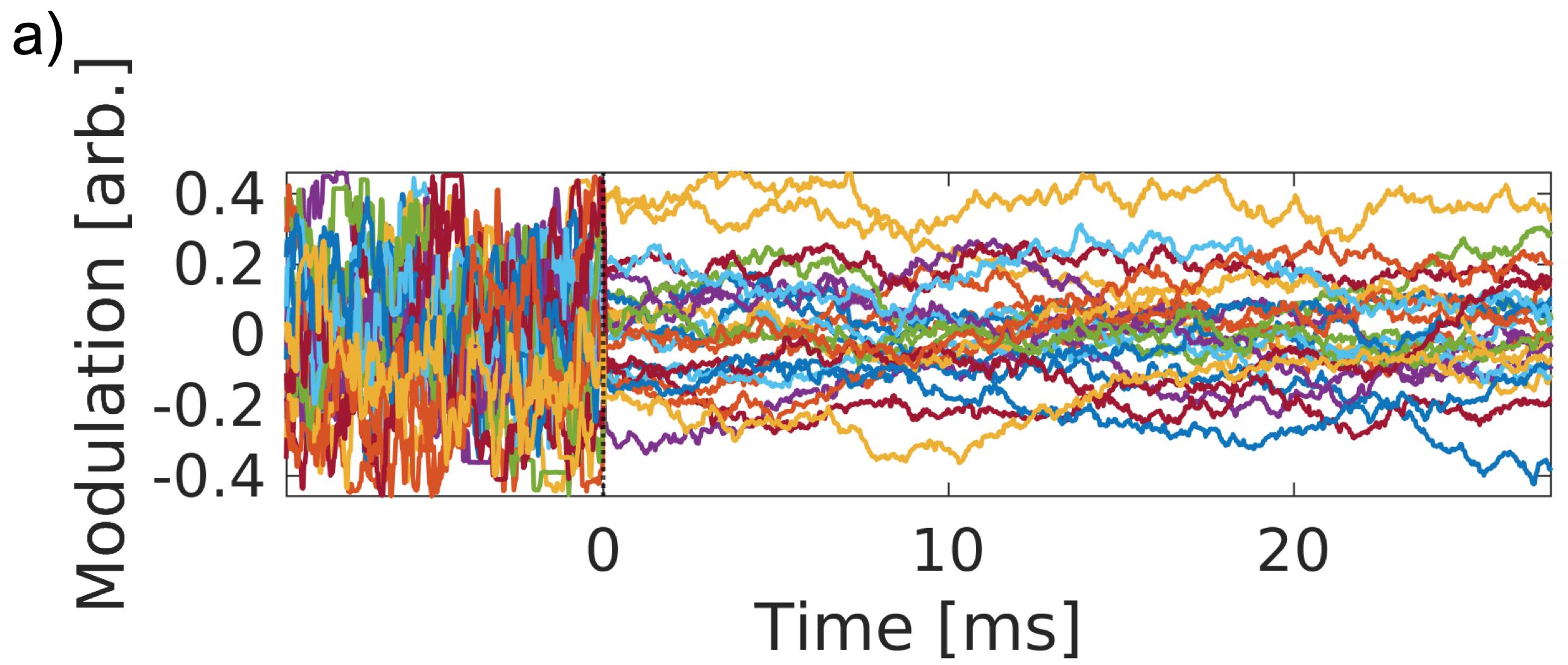}
		\label{fig:Modulation}}\\
	\centering{
		\includegraphics[width=0.35\textwidth]{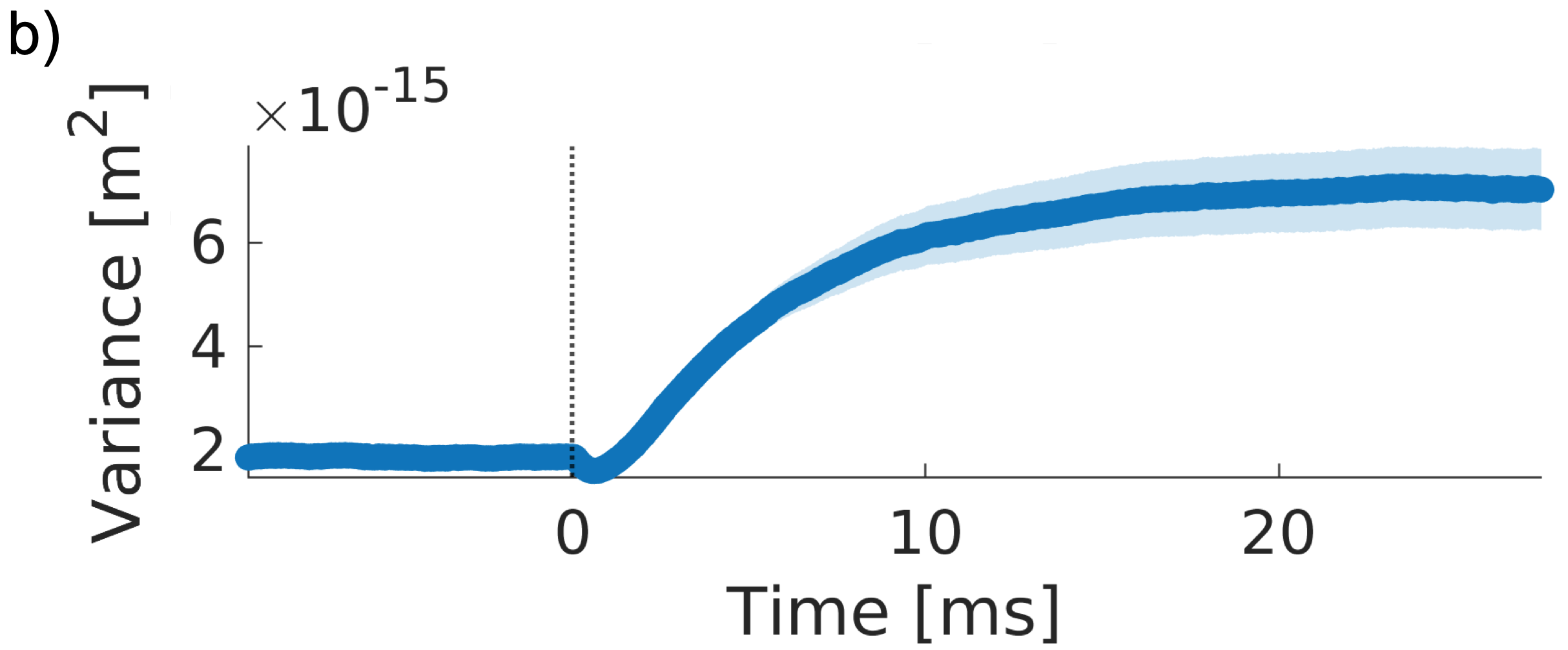}
		\label{fig:Variance}}\\
	\centering{
		\includegraphics[width=0.35\textwidth]{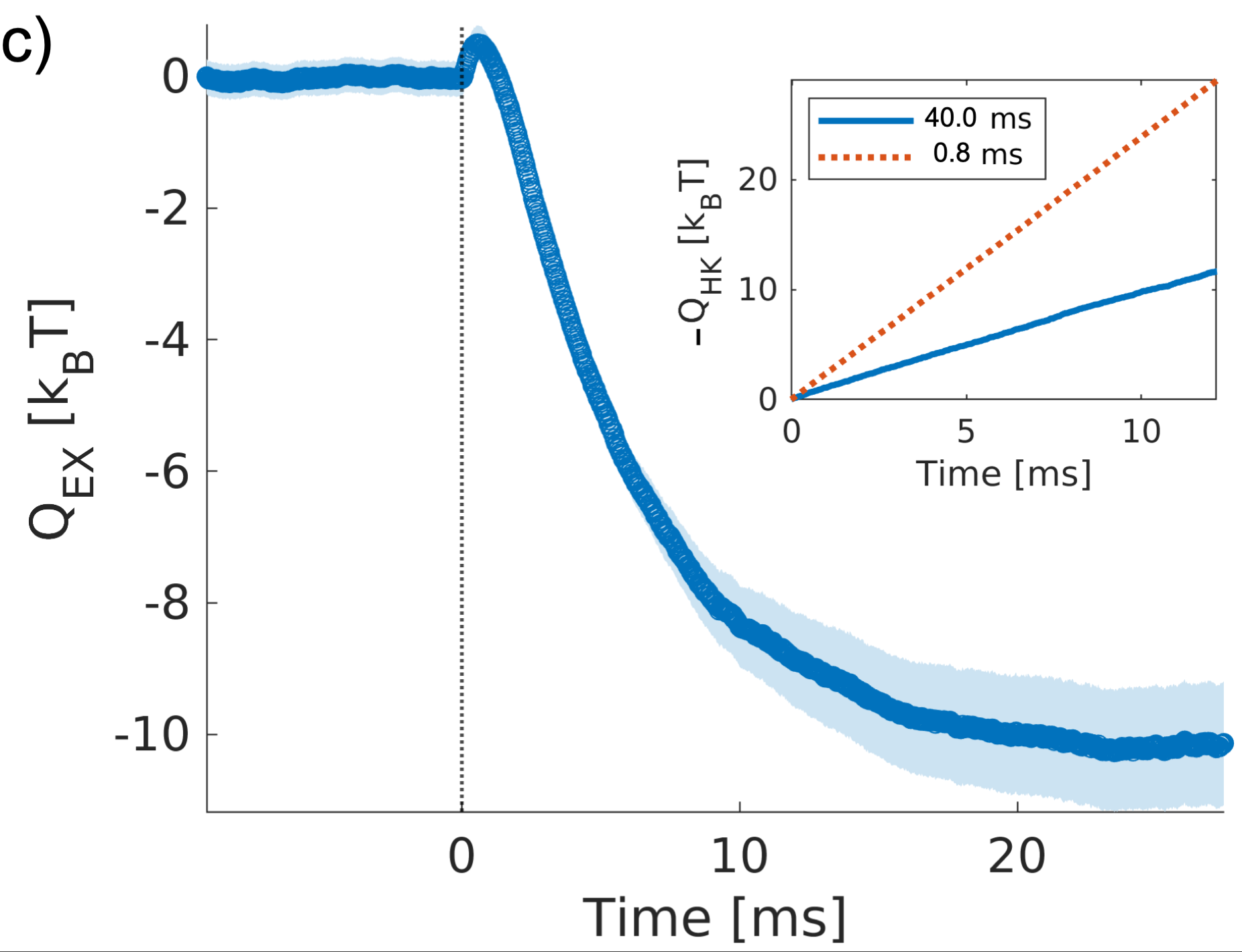}
		\label{fig:Heat1}}\\
	\caption{STEP protocol using the noise correlation time as control parameter. (a) Several digital realizations of the noise variable $\eta_t$, undergoing a change in the correlation time at $t=0$, from $\tau_c = 0.8\, \rm ms$ before the STEP to $\tau_c = 40 \,\rm ms$ after, while $\omega_0^{-1} = 2.1$ ms.
		(b) Corresponding experimental ensemble variance $\langle x_t^2\rangle$ of the positions of the bead; the shaded area represents the $99.7\%$ confidence interval.
		(c) Main plot: Cumulative excess heat in units of $k_B T$. Inset: Cumulative housekeeping heat before the STEP (blue line) and after the STEP (red dotted line). In the inset, the time origin is set in steady state for each regime. Sign conventions for the heat terms are used in agreement with \cite{Sekimoto1998}.
	}
\end{figure}
\vspace{1mm}

In Fig. 3, we show the results of a typical STEP protocol.
Figure \ref{fig:Modulation}(a) shows some realizations of the noise variable $\eta_t$ where the change in $\tau_c$ is visible at $t=0$.
In  Fig. \ref{fig:Variance}(b), we represent the variance of the position $x_t$ of the bead, which undergoes a threefold increase when the correlation time is changed.
The small dip right after the STEP is due to the fact that the non-Markovian trajectory is an integral of the noise $\eta_t$, hence needs a finite time to probe the total amplitude of $\eta_t$, as discussed in detail in the Appendix \ref{APPENDIX:VarStep}.

\highlight{
Even though the increase in variance in Fig.  \ref{fig:Modulation}(b) looks similar to the one induced by a rise in temperature, we emphasize that this process cannot be modeled through a mere effective temperature change.
Indeed, the increase in variance is only one of the many outcomes of the color protocol, which actually modifies all out-of-equilibrium properties, from the breaking of equipartition to the fluctuation-dissipation relations. These modifications cannot be understood simply as the consequence of an effective temperature change.}

\highlight{
Furthermore, an important difference with respect to temperature protocols is that, in our case, we do not change the amplitude of the \highlight{driving} noise, but only act on its spectrum by modifying the correlation time.}
In this sense, the protocol can seem costless from the experimentalist's point of view, as no additional power has to be provided to the  laser source.
As a comparison, we estimated the  equivalent power needed to induce the same increase in variance as in Fig. \ref{fig:Variance}(b) through a change in the noise amplitude, i.e., by changing \highlight{only} the diffusivity $D_a$. The result is that one would  need a laser intensity of $70~\rm mW$, whereas we used only $36~\rm mW$ in our color-based protocol.

From a thermodynamic point of view, our color-based protocol necessarily produces heat, which is then released in the thermal bath.
Following Sekimoto's treatment \cite{Sekimoto1998, SekimotoBook}, which was recently
applied to active matter \cite{Eichhorn2019}, the cumulative stochastic heat can be written (see Appendix \ref{APPENDIX:Thermo}):
\be
Q(t) = Q_{EX} + Q_{HK}= \frac{1}{2}\int_0^t\kappa \frac{dx^2 }{dt'}dt' - \gamma \int_0^t \sqrt{2D_a}\eta_{t'} \dot{x}_{t'} dt'.
\ee
The first term accounts for the \textit{excess} heat released during a transient evolution of the distribution, and vanishes for steady states \cite{Seifert2012}.
It is shown on the main graph of Fig. \ref{fig:Heat1}(c) with an energy release of $\approx 10 k_B T$.
The second term, expressed in terms of the cross-correlation $\langle \eta \dot{x}\rangle$, can be evaluated analytically by injecting $\dot{x}$ as from Eq. \eqref{eq:model} and is shown to grow linearly in time at steady state (see Appendix \ref{APPENDIX:Thermo}). This is the \textit{housekeeping} heat, representing the constant expense needed to maintain the system in its NESS.
It is displayed in the inset of Fig. \ref{fig:Heat1}(c) both before and after the protocol, showing that changing the correlation time affects the heat dissipation rate. Interestingly, the powers at play in our optical trap ($dQ_{HK}/dt  \approx 1 \,k_B T / \rm ms$)  are close to those involved in biological processes such as kinesin axonal transport \cite{Woehlke2000,Ariga2018,Ariga2021}.

\section{Harvesting information from the noise}
The protocol described in the above paragraphs seems to raise a paradox: after the transition, the heat released during the process appears to increase (see Fig. \ref{fig:Heat1}), while no further energy was injected in the system since only the spectrum of the noise was changed and not its amplitude. As the first principle of thermodynamics is of course not violated, this means that the coupling between the correlated noise bath and the bead has increased, so that energy can be transferred more efficiently from the former to the latter.

Nevertheless, it is illuminating to analyze this situation from an informational point of view. Indeed, a correlated noise carries more information and has lower entropy than a white, or a less correlated, noise. \highlight{In the framework of our experiments, this} information content can be measured by the spectral entropy $H_S$, which is just  the Shannon entropy in the frequency domain \cite{Zaccarelli2013}:
\be
H_S = - \sum_{i=1}^N P(\omega_i) \ln P(\omega_i) ,
\ee
where $P(\omega_i)$ denotes the normalized power spectral density of the signal $\eta$ at frequency $\omega_i$. The spectral entropy vanishes for a monochromatic signal and reaches its maximum $k_B \ln N$ for white noise.
Any correlated noise has an intermediate value of $H_S$.

\begin{figure}[htb]
	\centering{
		\includegraphics[width=0.8\linewidth]{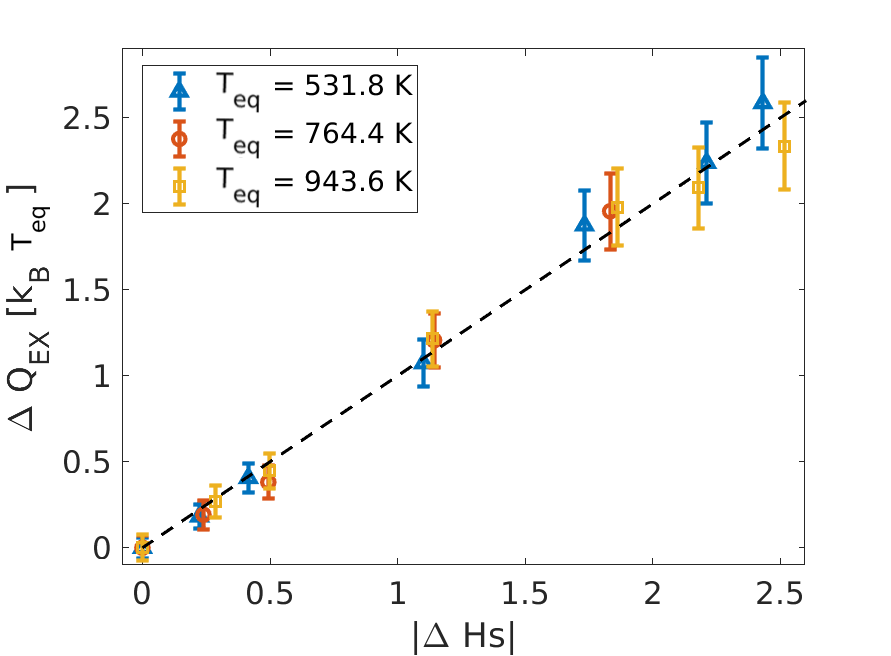}}
	\caption{Measured excess heat between two NESS \modif{(in units of $k_B T_{\rm eq}$) plotted as a function of the calculated spectral entropy, for three values of the noise amplitude, each characterized by an equipartition temperature $T_{\rm eq}$.} The dashed line has a slope equal to unity. Experimental details are given in Appendix \ref{APPENDIX_Hs}.
	}
	\label{fig:SpectralEntropyLin}
\end{figure}
\vspace{-1mm}

We measured $H_S$ for various correlation times and noise amplitudes.
\modif{Each series is labeled by the equipartition temperatures $T_{\rm eq}$, measured for a white noise of same amplitude.
For each temperature, we define $\Delta H_S = H_S(\tau_c) - H_S(\tau_{ref})$, and similarly for $\Delta Q_{EX}$, where $\tau_{ref}=0.5\,\rm ms$ is taken as a reference correlation time.}
Hence,  $\Delta H_S$ and $\Delta Q_{EX}$  represent respectively the informational expenditure and the corresponding energetic cost to go from the reference case to the colored case with correlation time $\tau_c$ through the STEP protocol described above.

\modif{The data obtained from several measurements are plotted in Fig. \ref{fig:SpectralEntropyLin} and nicely obey the relation $\Delta Q_{EX} / k_B  T_{\rm eq} = \Delta H_S$. This strikingly reveals that the excess heat produced in the process corresponds exactly, in units of $k_B  T_{\rm eq}$, to the injected information.}
\modif{The relationship was tested for exponentially-correlated noises of different amplitudes and over a wide range of correlation times, and it holds very neatly without any fitting or adjusting parameters. Although its extension to other classes of noises is less clear, exponentially-correlated noises constitute by far the most widely used and studied noises in stochastic thermodynamics}.
We stress that this expression is highly nontrivial: the left-hand side is a thermodynamic quantity related to the diffusive motion of the trapped bead, while the right-hand side captures the informational content of the colored bath generated by the laser beam. This remarkable result highlights the informational nature of our process, resolving what could appear as a paradoxically costless protocol.

\section{Conclusion}
We studied experimentally the diffusive motion of an optically trapped particle subjected to both a white thermal noise due to the surrounding fluid and a correlated noise generated by a digitally controlled fluctuating radiation pressure. This configuration constitutes a very accurate, controllable model of active biological matter.
\modif{Although it was implemented experimentally here using a microsphere trapped in a harmonic potential, and for the class of exponentially correlated noises, it can be easily extended to other types of tracers, potentials, and noises.}

Three major results were obtained: (i) thanks to the flexibility of our setup, we could explore an unprecedented range of regimes, most notably those characterized by long correlation times and strong amplitudes, which are unattainable in experiments with real active matter; (ii) by using the correlation time as a control parameter, we devised a protocol that drives the system from one NESS to another, at zero nominal energetic cost; (iii) we showed that the excess heat released during such a protocol is proportional to the spectral entropy of the \modif{exponentially correlated} noise, a relationship that is akin to Landauer's principle \cite{Plenio2001,Lutz2015}. The protocol harvests information from the colored noise, turns it into heat necessary for the transition between the two NESS, and finally releases it to the surrounding environment.
The ubiquity of non-equilibrium steady states in biological systems, including changes in the spectrum of the bath through time (for example, during mitosis \cite{Hurst2021, Taubenberger2020}), \highlight {but also in biochemical sensing systems \cite{Loos2020, Lan2012} or within active crowded environments \cite{VolpeReview2016} such as microtubules assemblies \cite{Hess2017}, suggests exciting applications for the present findings. In particular, we anticipate that the relation between heat and spectral entropy could serve as a new tool for the study of non-equilibrium systems in non-trivial baths.}

\section*{Acknowledgments}

We thank Samuel Albert, Minghao Li, and Laurent Mertz for discussions. This work is part of the Interdisciplinary Thematic Institute QMat of the University of Strasbourg, CNRS and Inserm. It was supported by the following programs: IdEx Unistra (ANR-10-IDEX- 0002), SFRI STRATUS project (ANR-20-SFRI-0012), ANR Equipex Union (ANR-10-EQPX-52-01), the Labex NIE project ANR-11-LABX-0058-NIE, and USIAS (ANR-10-IDEX- 0002-02), under the framework of the French Investments for the Future Program. \modif{We thank an anonymous referee whose questions inspired us to extend our analysis on heat and information. }

\appendix

\section{Experimental setup and calibration}
\label{APPENDIX_samples}

\begin{figure}[htb]
  \centering{
    \includegraphics[width=0.6\linewidth]{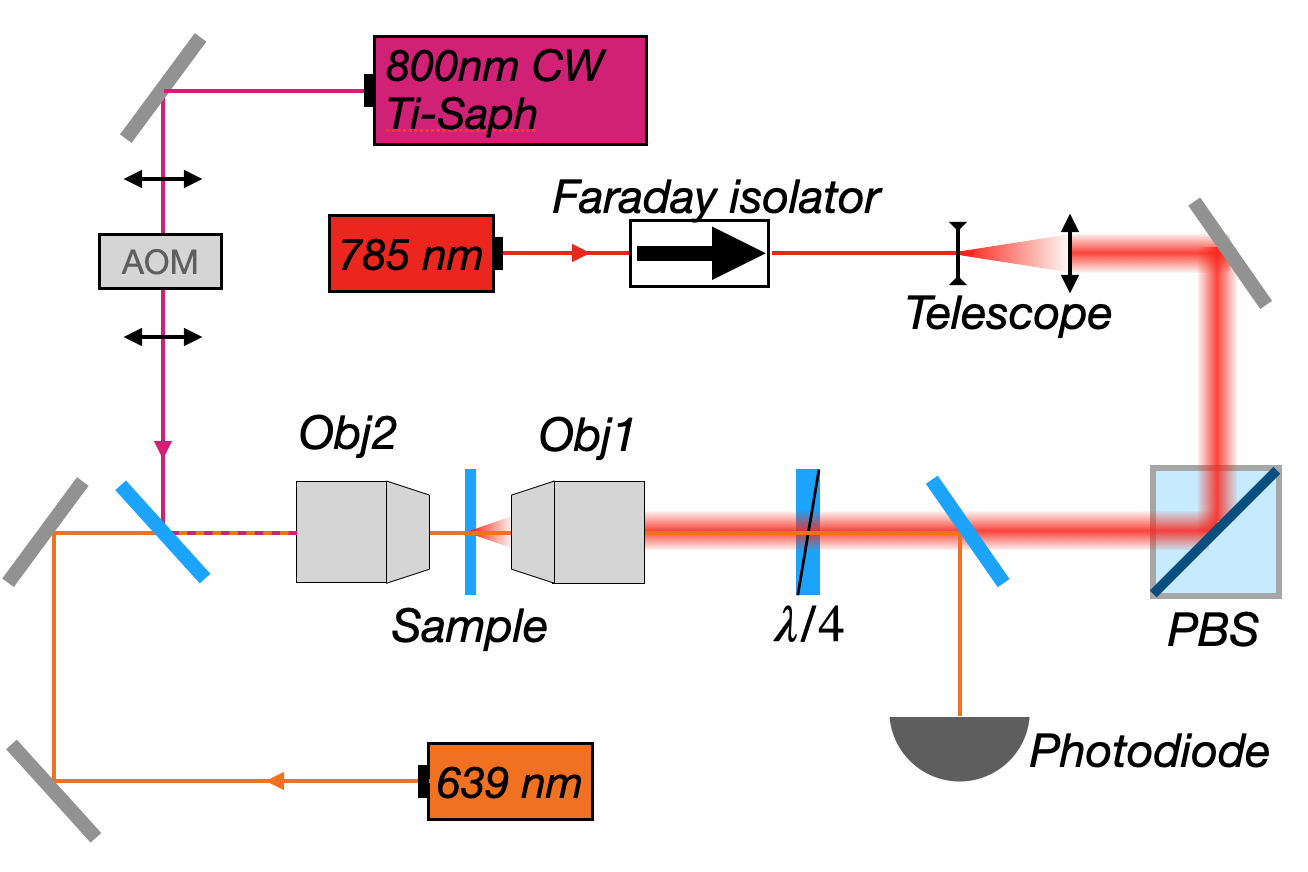}}
  \caption{{
  		Simplified view of the optical trapping setup. The sphere is suspended in water inside the \textit{Sample} cell inserted between the two objectives Obj1 and Obj2. The $785~\si{\nano\meter}$ trapping beam is drawn in red, the $800~\si{\nano\meter}$ radiation pressure beam in purple. The intensity of this beam controlled by the acousto-optic modulator (AOM). The instantaneous position of the trapped bead is probed using the auxiliary $639~\si{\nano\meter}$ laser beam, drawn in orange, whose scattered signal is sent to a high-frequency photodiode.
  }}
  \label{fig:schema}
\end{figure}

Our experimental setup consists in optically trapping, in a harmonic potential, a single dielectric bead ($3~\si{\micro \meter}$ polystyrene sphere) in a fluidic cell filled with dionized water at room temperature.
The harmonic potential is induced by focusing inside the cell a linearly polarized Gaussian beam ($785~\si{\nano\meter}$, CW $110$ mW laser diode, Coherent OBIS) through a high numerical aperture objective (Nikon Plan Apo VC, $60\times$, NA$=1.20$ water immersion, Obj1 on Fig. \ref{fig:schema}). An additional force in the form of radiation pressure is applied to the sphere using an additional high-power laser ($800~\si{\nano\meter}$, CW $5$ W Ti:Sa laser, Spectra Physics 3900S).
The intensity of this radiation pressure beam is controlled by an acousto-optic modulator (Gooch and Housego 3200s, AOM on Fig. \ref{fig:schema}) using a digital-to-analogue card (NI PXIe 6361) and a \textsc{python} code.

The instantaneous position $x_t$ of the sphere along the optical axis is measured by recording the light scattered off the sphere of a low-power $639~\si{\nano\meter}$ laser (CW $30$ mW laser diode, Thorlabs HL6323MG), sent on the bead via a second objective (Nikon Plan Fluor Extra Large Working Distance, $60\times$, NA$=0.7$, Obj2 on the figure). The scattered light is collected by Obj1 and recorded by a photodiode ($100$ MHz, Thorlabs Det10A). The recorded signal (in V/s) is amplified using a low noise amplifier (SR560, Stanford Research) and then acquired by an analog-to-digital card (NI PCI-6251). The signal is filtered through a $0.3$ Hz high-pass filter at 6 dB/oct to remove the DC component and through a $100$ kHz low-pas filter at 6 dB/oct to prevent from aliasing. The scattered intensity varies linearly with the position of the trapped bead $x_t$ for small enough displacements and we make sure to work in the linear response regime of the photodiode so that the recorded signal is linear with the intensity, resulting in a voltage trace well linear with $x(t)$.

\begin{figure}[htb!]
	\centering{
		\includegraphics[width=0.4\linewidth]{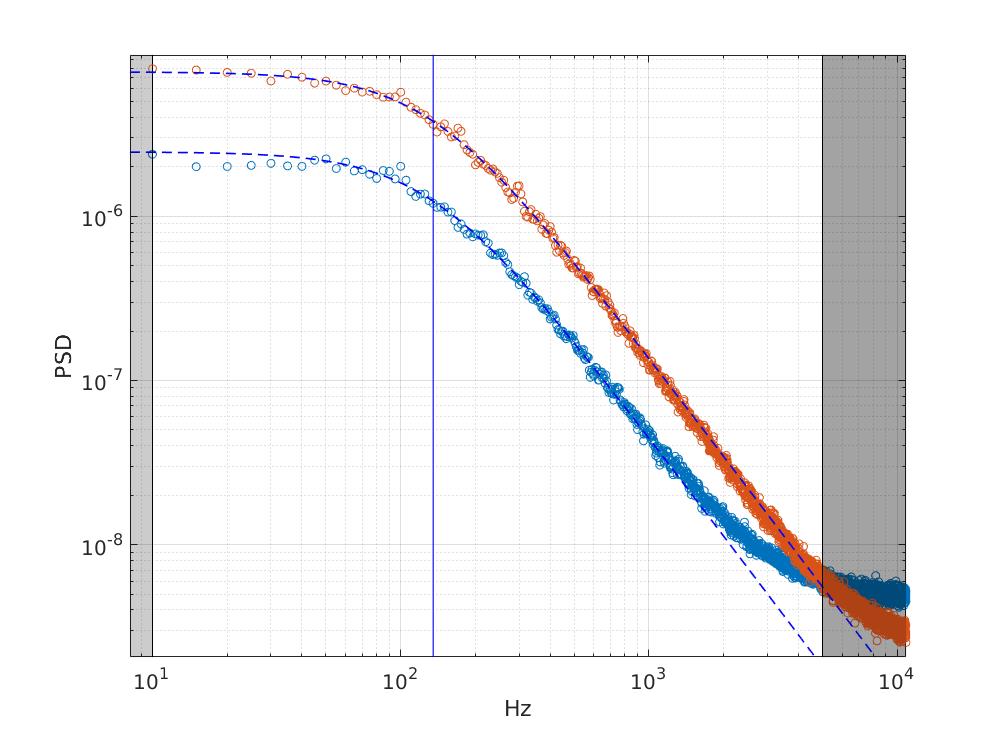}}
	\caption{{
			Power spectral density $S_x[\omega]$ of the bead position, recorded with no additional noise, i.e. solely driven by thermal fluctuations (blue circles) and with additional white noise injected inside the trap (red circles) using the auxiliary radiation pressure laser made noisy via the AOM. In both cases, the dashed lines correspond to Lorentzian fits, with shaded regions indicate the limits of the fits. The vertical line mark the position of the trap roll-off frequency at $\sim10^2 \si{\hertz}$. Note the onset of the electronic noise floor at high frequencies.
	}}
	\label{fig:Calibration}
\end{figure}
\vspace{3mm}

The calibration of the recorded voltage is done by fitting the motional Lorentzian spectrum of the sphere -see Fig. \ref{fig:Calibration}- from which is extracted a calibration coefficient expressed in $\si{\meter}/\si{\volt}$, generally $\sim 10^{-7} \si{\meter/\volt}$. The noise added by the radiation pressure laser modifies the dynamics of the bead, but without changing the properties of the trapping potential. As presented in Fig. \ref{fig:Calibration}, we carefully verify that an added white noise that mimics a higher temperature, only leads to an increase in the power spectral density amplitude (as expected when increasing the kinetic temperature of the bead) without modifying neither its Lorentzian profile nor the roll-off frequency of the trap, left unchanged at $\approx 150 \si{\hertz}$.

\section{External radiation pressure force acting as a bath: from noise generation to active protocols}
\label{APPENDIX_ExtNoise}

\begin{figure}[htb]
	\centering{
		\includegraphics[width=0.6\linewidth]{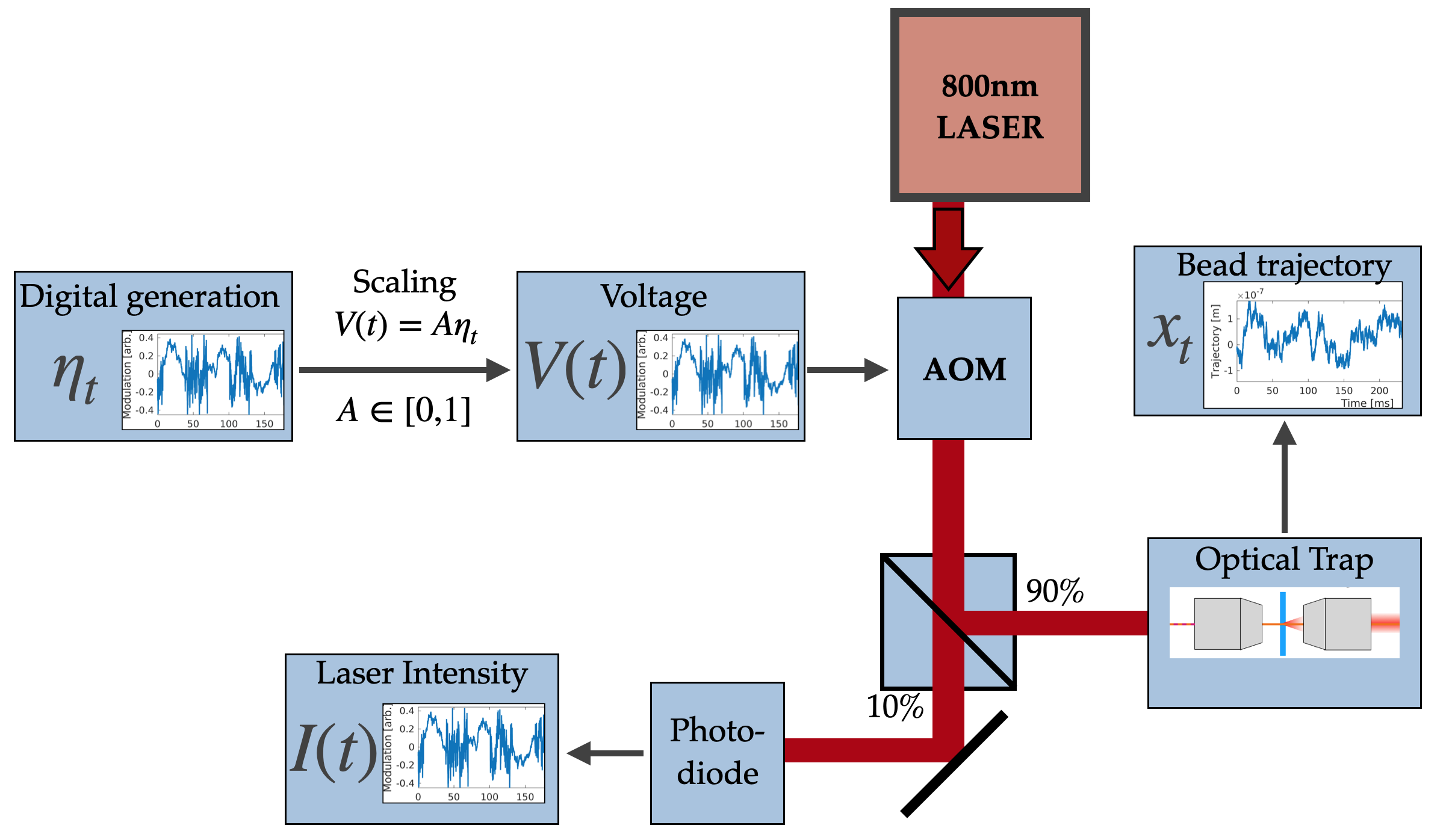}}
	\caption{{Schematic representation of the signal and data acquisition processing implemented in our experiments. The noise $\eta_t$ is digitally generated, scaled to a voltage $V(t)$ that can be sent to the acouto-optic modulator (AOM), producing a diffracted beam whose intensity varies linearly with this input voltage. The laser beam diffracted through the AOM exerts a radiation pressure on the optically trapped sphere, whose position is recorded as detailed in Sec. \ref{APPENDIX_samples}. A small part of the laser beam is also measured and used to monitor and evaluate the noise $\eta_t$ as it enters the trap.
	}}
	\label{fig:SchemaLogic}
\end{figure}

We generate the external noise following the sequence described in Fig. \ref{fig:SchemaLogic}. Using a \textsc{python} code with a build-in random noise generator, we can easily generate a white noise.
\highlight{In the case of white noise, each random number is independent, hence due to the law of large number (more precisely to the central limit theorem), it will yield a Gaussian density for the positions $x_t$.
The choice of a white noise probability density is therefore not important. To maximize the available dynamical range, we thus simply use a uniformly distributed noise. This free choice for the noise probability density will no longer be possible for correlated noise. We will use a Gaussian probability density in that case.}

\vspace{3mm}

In contrast, for a colored noise, both correlation and distribution matter. In this case, we use a Gaussian exponentially correlated noise, given as the solution of the Ornstein-Uhlenbeck process
\be
d\eta_t = -\omega_c \eta_t dt + \sqrt{2 \alpha \omega_c}dW_t,
\label{eq:noise}
\ee
where $\omega_c$ is the inverse characteristic time of the noise and $\sqrt\alpha$ its amplitude.
The variance of such a process is $\langle \eta_t ^2 \rangle = \alpha$.
This correlated noise will then enter into the Langevin equation as an external random force field
\be
	\gamma \dot{x}_t = -\kappa x_t + \gamma \sqrt{2 D} \xi_t + F_{ext}(t)
\ee
with $D = k_B T/\gamma$ the diffusion coefficient in the thermal bath.

\begin{figure}[htb]
	\centering{
		\includegraphics[width=0.4\linewidth]{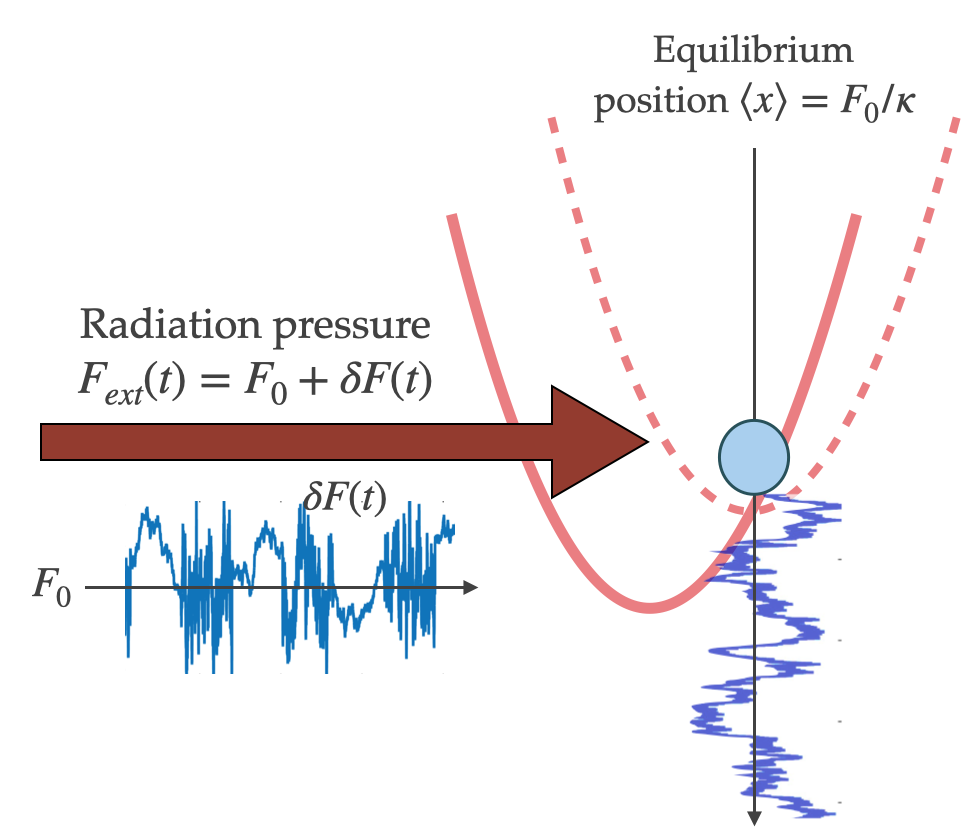}}
	\caption{\highlight{Schematics of the action of the radiation pressure force centered around a mean value $F_0$, which induces a small displacement of the mean position $\langle x_t \rangle$ of the trapped microsphere. This mean shift can be filtered away when recording the instantaneous position of the sphere, so that the remaining force corresponds the fluctuating part $\delta F(t)$ only. Combined with the restoring force $-\kappa x_t$, it can give both negative (pulling) and positive (pushing) values.
	}}
	\label{fig:SchemaRadPres}
\end{figure}
\vspace{3mm}

The external radiation pressure force $F_{ext} = F_0 + \delta F(t)$ is centred around a mean value $\langle F_{ext} \rangle = F_0$ and with a zero mean noise part $\langle \delta F(t) \rangle = 0$.
\highlight{The effect of the radiation pressure external force is schematized on Fig. \ref{fig:SchemaRadPres}, where the mean displacement of the microsphere in the trap induced by the mean radiation pressure value is shown, together with the superimposed action of the fluctuating bath around this mean value.}
The average term $F_0$ vanishes trivially  when looking at the centred process $x_t - \langle x_t \rangle = x_t - F_0/\kappa$, which is always the case in our experiments.
Random, the external force acts as a secondary bath. It can thus be recast as $F_{ext}(t) = \gamma \sqrt{2 D_a} \eta_t$, \textit{i.e.} on the same footing as the thermal force $F_{th}(t) = \gamma\sqrt{2D}\xi_t$ where $D$ is in $\si{\square\meter/\second}$ and $\xi$ (the time derivative of a Wiener process) is in $\si{\sqrt{\hertz}}$.
This leads us to introduce an \textit{active} diffusion coefficient $D_a$ having the same dimension as $D$, associated with the noise $\eta_t$ solution of Eq. (\ref{eq:noise}). This gives the noise of dimension $[\eta_t] \equiv [\sqrt\alpha] \equiv \si{\sqrt{\hertz}}$, just like the thermal noise term $\xi_t$.

An important asset of our work is the flexibility of our scheme for controlling the color and the amplitude of the noise, which demands to keep $\omega_c$ and $\alpha$ independent. To do so, we first generate numerically with the \textsc{python} code, under a sampling frequency of $20\si{\kilo\hertz}$, the noise $\eta_t$ of variance $\alpha$, scaled to the desired amplitude $\in [0, 1]$ which corresponds, in Volts, to the range fixed by the radio-frequency generator driving the acousto-optic modulator (AOM).
The AOM response is calibrated to yield a linear relation between the input voltage and output laser intensity in the first order diffracted beam.
The numerically generated noise is thereby encoded into a radiation-pressure laser intensity noise sent to the bead. This noise intensity acting on the bead depends both on the gain of the AOM-diffracted beam and on the choice of the radiation-pressure laser intensity.

In our experiments, the amplitude of the noise is independent of its color, which is different from the choice made in \cite{DiLeonardo2014}, where $\alpha$ scales as the square root of the inverse correlation time of the noise. This choice necessarily induces the interplay between both correlation times and amplitudes that we want to avoid.
The actual amplitude of the noise experienced by the bead, which depends not only on the radiation-pressure laser intensity but also on the optomechanical coupling between this laser beam and the trapped sphere, will be taken into account in $F_{ext}$ via the active diffusion coefficient $D_a$. This implies that choosing $\alpha = 1$ is the simplest option.  However, we keep the $\alpha$ term for clarity, as a purely dimensional constant.

\begin{figure}[htb!]
	\centering{
		\includegraphics[width=0.35\textwidth]{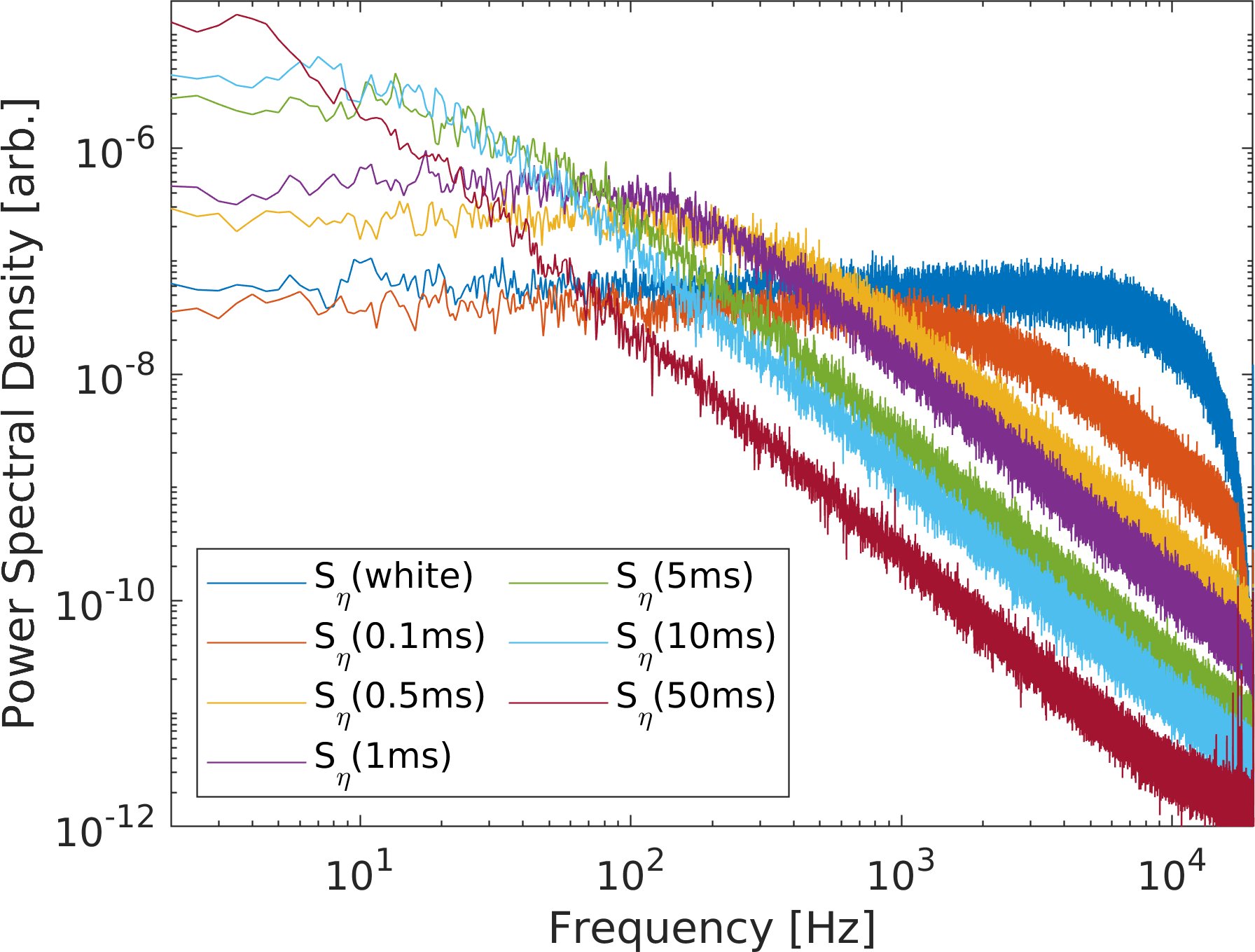}
		\label{fig:AllSmoothSpectra}}
	\centering{
		\includegraphics[width=0.35\textwidth]{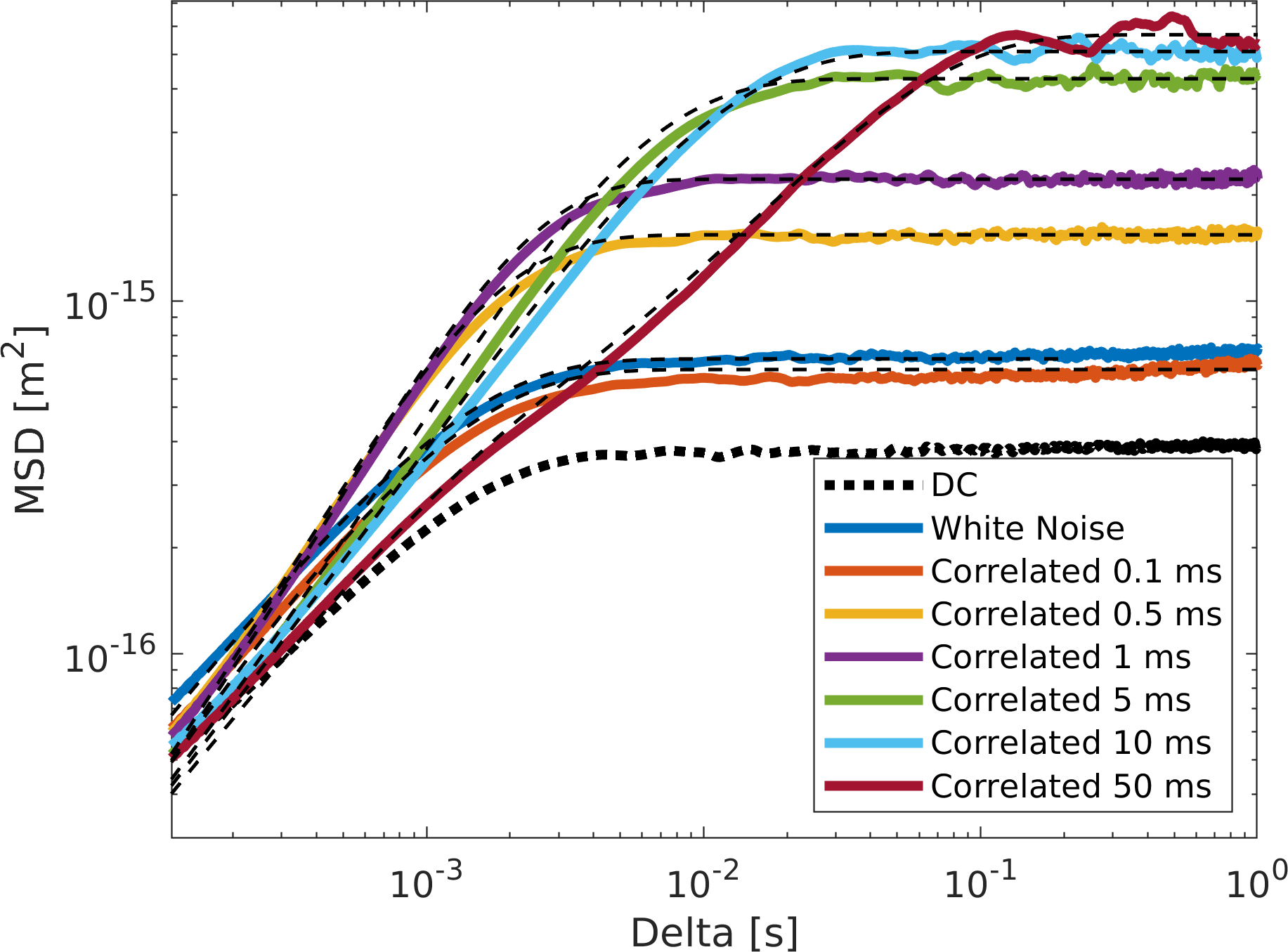}
		\label{fig:SomeMSDWithoutFit}}
	\caption{{ (a) Power spectral density associated with different noises $\eta_t$ (for each colored curves, the  corresponding correlation time is indicated within brackets in the legend). These spectra are measured directly from the laser intensity signal that is sent on the trapped bead as a radiation pressure.
			(b) Mean squared displacement of the sphere for each of the noises presented in panel (a). The DC case corresponds to the absence of additional noise, with $F_{ext} = F_0$. We observe that the white noise case and the first colored noise case (correlation time $0.1\si{\milli\second}$) are almost identical. Superimposing the fit performed with the analytical expression for the mean squared displacement (see below, Sec. D) enables one to extract the active diffusion coefficients $D_a$ for each case.
	}}
\end{figure}
\vspace{3mm}

On Fig. \ref{fig:AllSmoothSpectra} (a), we show the power spectral densities of different noises, from white to colored.
At high frequencies ($>3000~\si{\hertz}$, see Fig. \ref{fig:Calibration}), the signal is dominated by the electronic noise of the experiment, limiting the spectral bandwidth of interest from a fraction of Hz to a few kHz.
The blue curve in Fig. \ref{fig:AllSmoothSpectra} corresponds to a white noise generated over the desired bandwidth where we see its flat spectrum covering all the response region of the bead, up to the $20~\si{\kilo\hertz}$ limit of the generation sampling frequency.
The other curves are the different colored noises, with correlation times spanning from $0.1$ to $50~\si{\milli\second}$.
On Fig. \ref{fig:SomeMSDWithoutFit} (b), we show the mean squared displacement (MSD) associated with each noise.
The black curve shows the DC case with $F_{ext} = F_0$ where no noise is added.
The blue curve corresponds to the white noise drive (blue spectrum of Fig. \ref{fig:AllSmoothSpectra} (a)), slightly above the thermal MSD as a consequence of the increase in effective temperature. The orange curve gives the first colored noise case (orange spectrum of Fig. \ref{fig:AllSmoothSpectra} (a)). We can note that the responses to a white noise and to a colored noise of correlation time $0.1 ~\si{\milli\second}$ are similar, this colored noise being "almost white". This implies that longer ($>0.1$ ms) correlation times are needed to make a clear difference between white and colored cases (as seen for the next colored noise with correlation time $0.5 ~\si{\milli\second}$). The other curves are the MSD corresponding to the different noises of Fig. \ref{fig:AllSmoothSpectra} (a).\\

The white noise is not generated here as a $\omega_c \rightarrow \infty$ limit.
Our choice for the correlation function $\langle \eta_t \eta_s \rangle = \alpha e^{-\omega_c |t-s|}$ with independent $\alpha$ and $\omega_c$ yields, as a consequence, that when $\omega_c \rightarrow \infty$, the amplitude vanishes over a finite bandwidth, since the total integral $\alpha$ is conserved.
It is clear on the different observables we measure -- such as PSD, MSD or heat -- that the limit $\omega_c \rightarrow \infty$ yields a vanishing additional noise, leading to normal diffusion at room temperature.
As explained above, the white noise with a finite amplitude over a finite bandwidth is generated with an independent method.
Yet, as it is also seen on Fig. \ref{fig:SomeMSDWithoutFit}, a colored noise with a finite $\omega_c$ can itself yield an identical response as that of a white noise.\\

\modif{
	As mentioned in the main text, when the noise applied is white, the MSD takes the simple form of a thermal Brownian motion MSD, but with a modified prefactor $2D_{\rm eq}/\omega_0 = 2k_B T_{\rm eq}/\kappa$, where $T_{\rm eq}$ corresponds to the temperature obtained through equipartition $T_{\rm eq} = \kappa \langle x^2 \rangle / k_B$.
	When the noise is correlated, the MSD is characterized by two independent coefficients $D$ and $D_a$ which cannot be combined as a single prefactor.
	The temperature obtained by equipartition $T_{\rm eq} = \kappa \langle x^2 \rangle / k_B$ depends on the correlation time of the noise and is not sufficient to characterize the MSD.
	In the following, we focus on the relation between the amplitude of the noise applied to the microsphere, the correlation time of the bath and the resulting $D_a$, $T_{\rm eq}$ and motional variance.
	We emphasize that $T_{\rm eq}$ has a clear and simple meaning when the driving noise is white, and will therefore be used as a way to characterize the amplitude of the driving noises.
	It will also often be used to normalize quantities of interest by a well controlled measure of the bath amplitude.}\\

\begin{figure}[htb]	
		\centering{
		\includegraphics[width=0.35\textwidth]{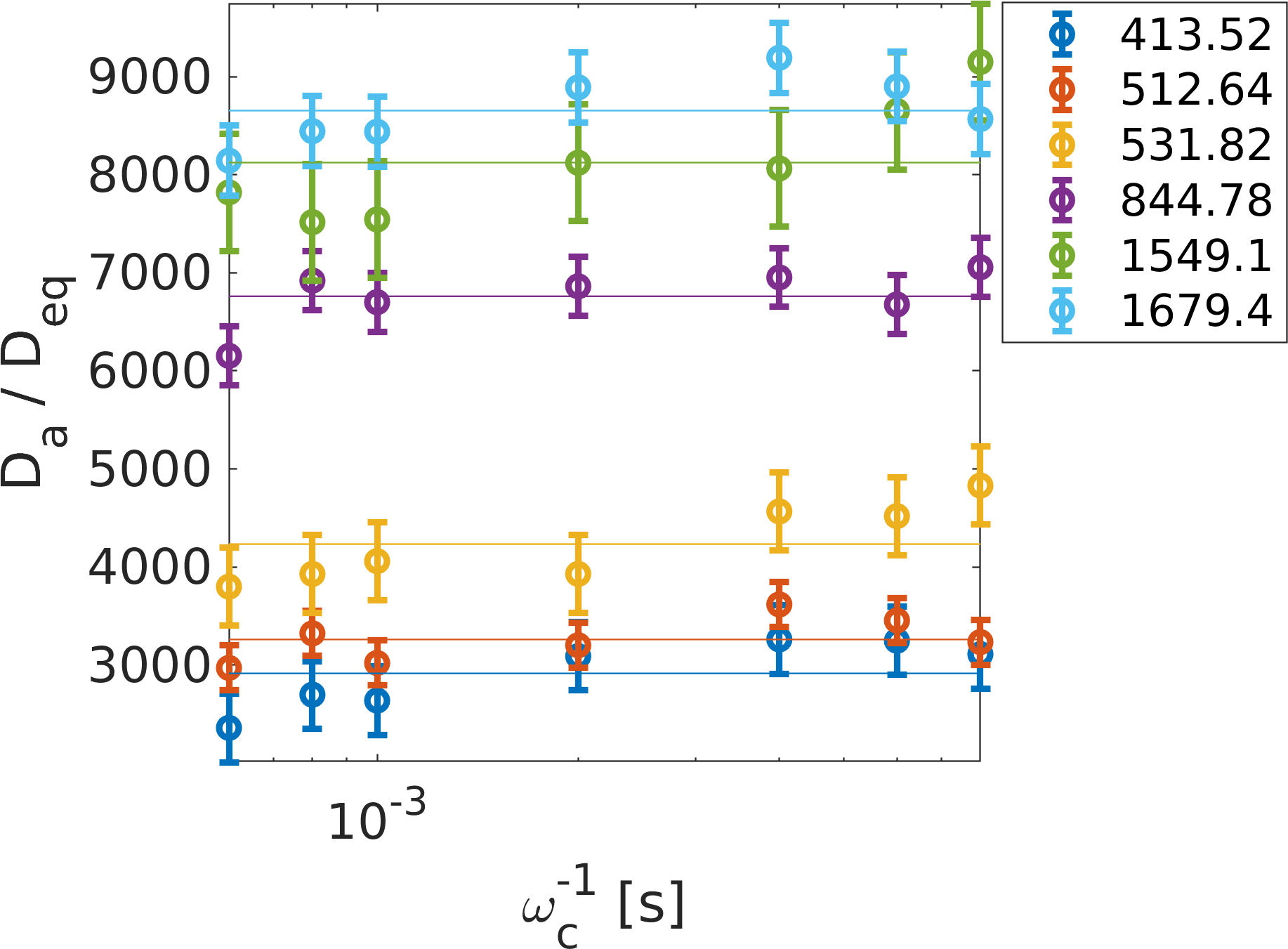}
		\label{fig:Da}}
	\centering{
		\includegraphics[width=0.35\textwidth]{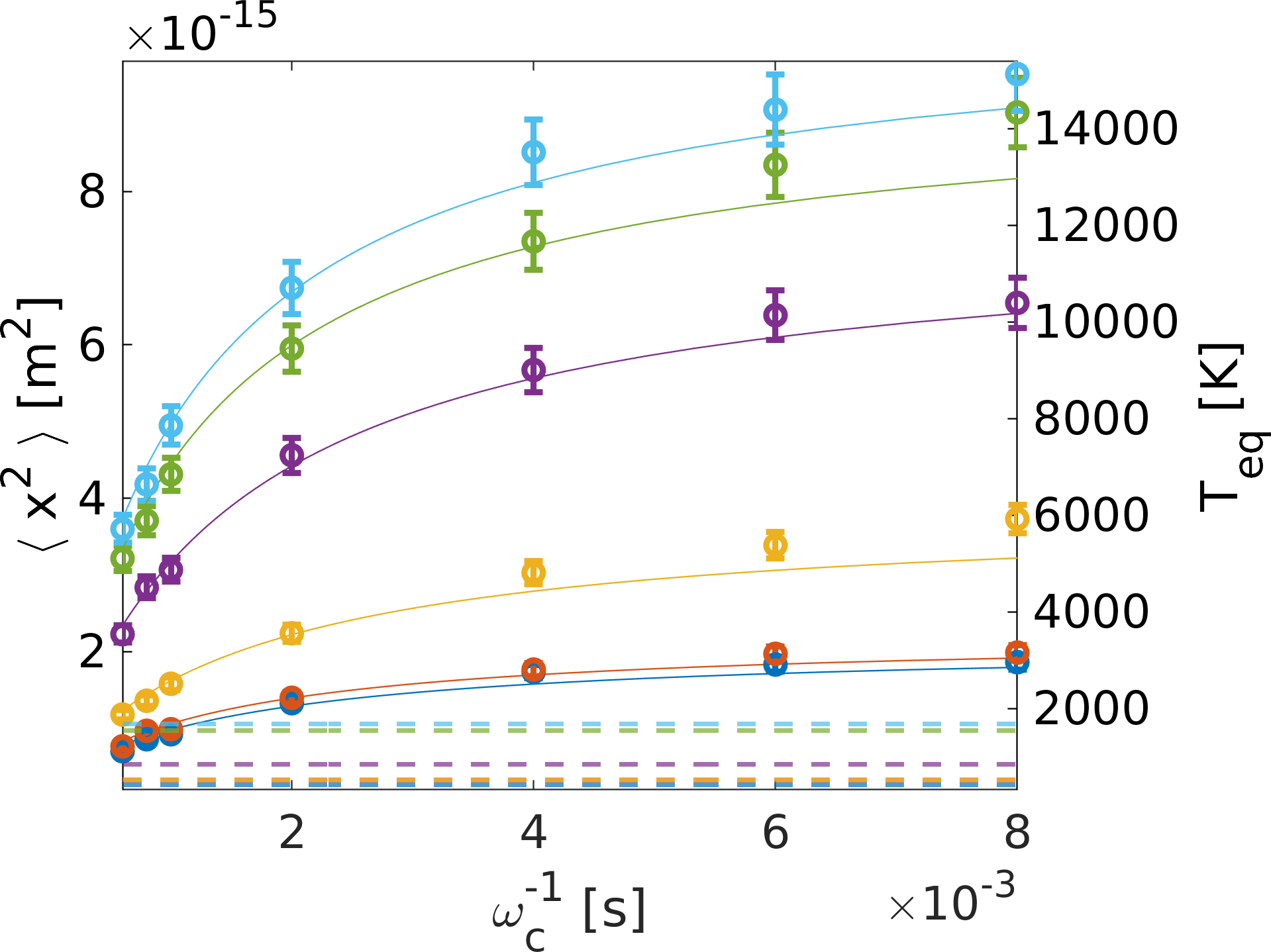}
		\label{fig:VarVsTauC}}
	\caption{{\modif{(left panel) Active diffusion coefficient $D_a$ normalized to the white noise equipartition diffusion coefficient $D_{\rm eq} = k_B T_{\rm eq}/\gamma$ as a function of the correlation time $\omega_c^{-1}$. This is plotted for various noise amplitudes, characterized by an equipartition temperature $ T_{\rm eq}$ computed for a white noise of same amplitude (in the legend, from $413.52$ to $1679.4$ Kelvin). (right panel) Motional variances $\langle x^2 \rangle$ (left axis) and related equipartition temperatures $ T_{\rm eq} = \kappa \langle x^2 \rangle / k_B$ (right axis) as a function of the correlation time $\omega_c^{-1}$ and for various noise amplitudes, characterized by the same white noise equipartition temperatures, also shown on the graph as constant dashed lines. Circles are experimental variances and solid lines represent the analytical result Eq. (\ref{eq:var}). Color coding is the same as in the left panel.
}
	}}
	\label{fig:DaVarVsTauC}
\end{figure}
\vspace{3mm}

\modif{
Figure \ref{fig:Da} (left panel) displays the constant active diffusion coefficient $D_a$ as a function of the scanned correlation times $\omega_c^{-1}$.
Each color, from deep blue to light blue corresponds to a different noise amplitude, characterized by the equipartition temperature $T_{\textrm{eq}}$ determined for a white noise of the \textit{same} amplitude (shown in the legend, ranging from $413.52$ K to $1679.4$ K).
The white noise equipartition temperatures $T_{\textrm{eq}}$ are used to normalize the active diffusion coefficient $D_a$ by $D_{\textrm{eq}} = k_B T_{\rm eq}/\gamma$.
Keeping the same amplitude $\langle \eta^2 \rangle$, we can scan the range of colored noise, by varying $\omega_c$.
For each noise amplitude, we can compute the motional variance $\langle x^2 \rangle$ as a function of the correlation time $\omega_c^{-1}$ and the related equipartition temperature $T_{\textrm{eq}} = \kappa \langle x^2 \rangle / k_B$.
We represent the motional variance and equipartition temperature on Fig. \ref{fig:DaVarVsTauC} (right panel).
Both quantities strongly depend on the correlation time and are well captured by the analytical expression Eq. (\ref{eq:var}).
The equipartition temperatures, reaching values above $10^4$ K, are always higher than the ones obtained with a white noise of same amplitude, which are displayed as dashed horizontal lines.}

\modif{For the analysis of the relation between heat and information, the excess heat is normalized by $k_B T_{\textrm{eq}}$ where $T_{\textrm{eq}}$ is evaluated with a white noise, providing a robust characterization of the bath energy.
}

\begin{figure}[htb!]
		\centering{
			\includegraphics[width=0.4\textwidth]{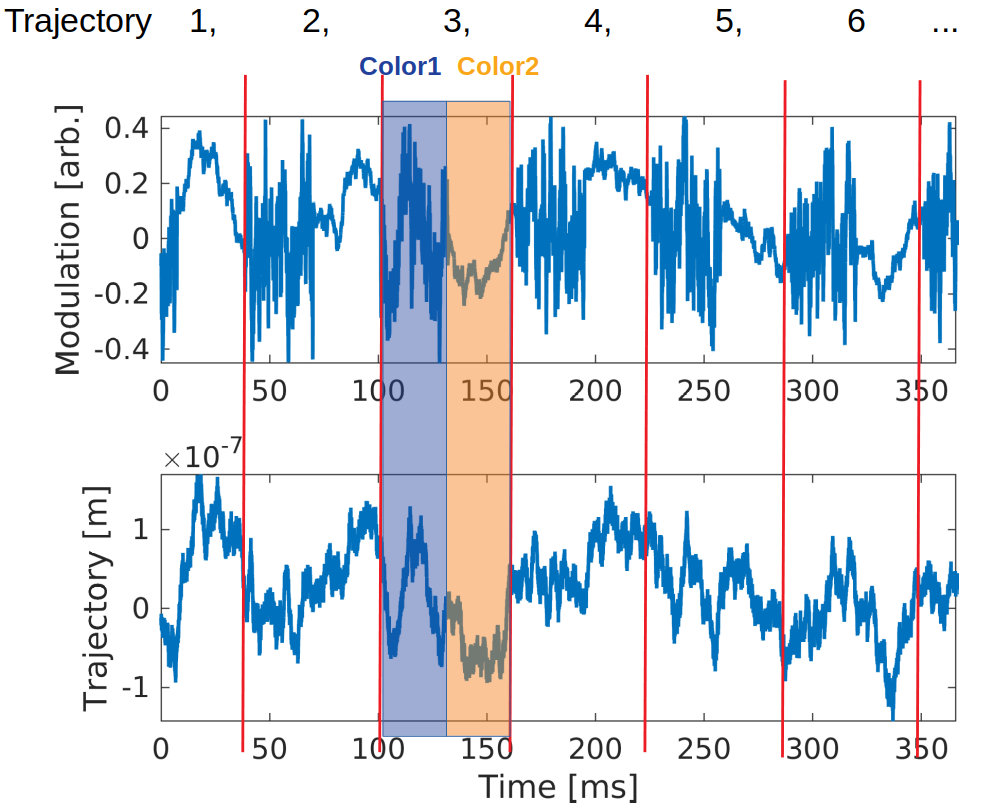}
			\label{fig:LongTraj}}
		\centering{
			\includegraphics[width=0.4\textwidth]{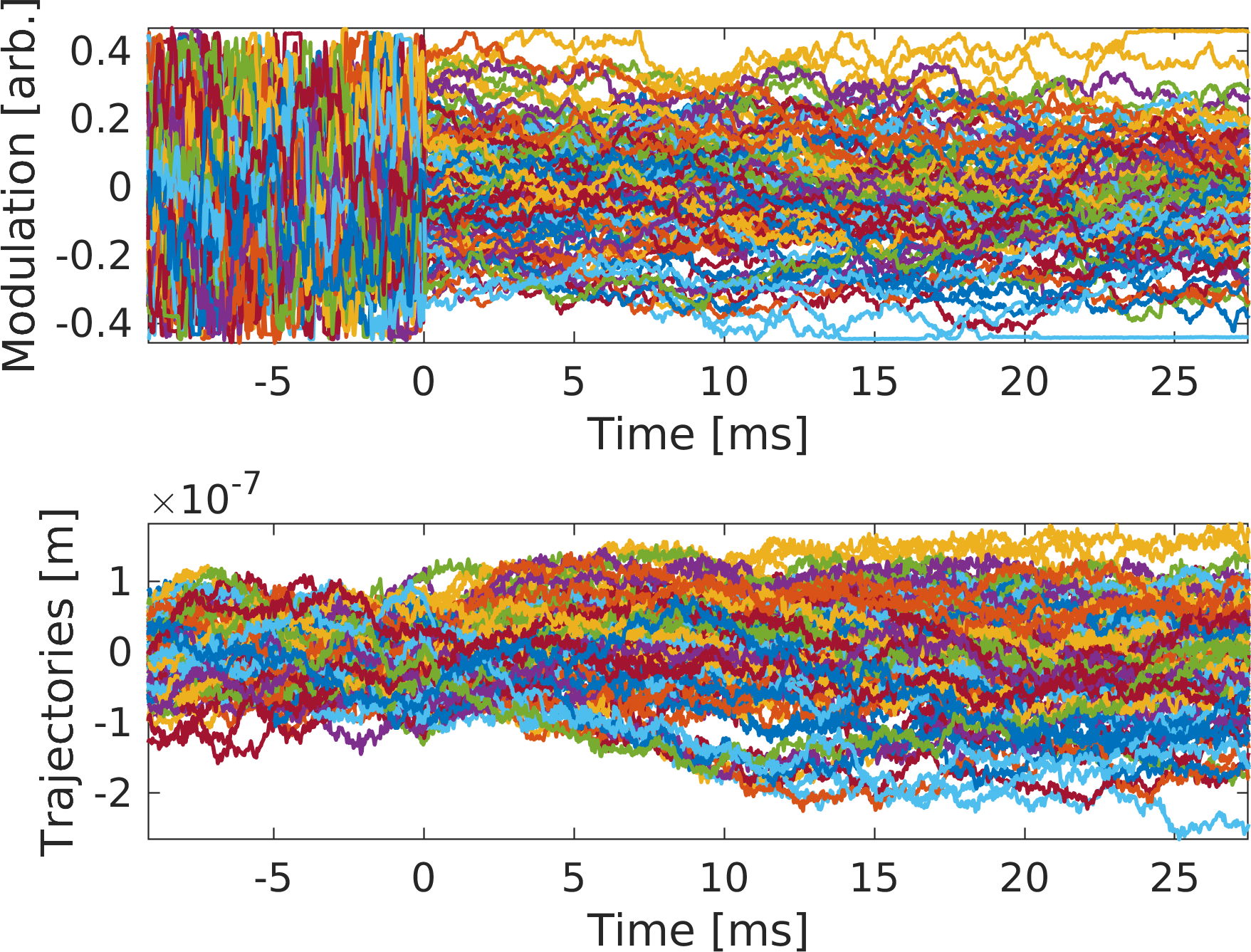}
			\label{fig:EnsembleNoiseAndTraj}}
		\caption{{ (a) Temporal noise series $\eta_t$ modulated between two correlation times (two colors) following a $20$ Hz square modulation function (top) and the resulting trajectory $x_t$ (bottom).
				(b) Since the time between switches is long, the long trajectory is cut and reshaped as an ensemble of independent synchronous trajectories. In the upper panel, the $\{\eta_t^i\}$ ensemble clearly displays at $t = 0$ the instantaneous change in correlation with constant amplitude. In the lower panel, the ensemble of trajectories $\{x_t^i\}$ of the bead show the progressive change in the motional variance that results from the step-like change of correlation times of the bath.
		}}
\end{figure}
\vspace{3mm}

To study protocols, we need an ensemble of independent trajectories all experiencing the same parameter changes.
In our experiments, this is a change in correlation time $\omega_c(t)$ that we modulate in a step-like way from an initial $\omega_c^i$ to a final $\omega_c^f$ values.
With one single bead in the optical trap, the ensemble is drawn out of a long time series, for which the ergodic hypothesis is crucial and was carefully checked as discussed in Sec. \ref{APPENDIX_Ergodicity}.
We produce one long noise sequence $\eta_t$ where a large number of correlation time changes are produced following a $\omega_c^i/\omega_c^f$ square modulation at a low enough repetition rate (a few tens of Hertz). This modulation sequence is sent to the bead via the radiation pressure laser.
The corresponding trajectory $x_t$ of the bead relaxes to one steady-state between each change on $\omega_c$. This long trajectory is cut and reshaped into an ensemble of trajectories $\{x_t^i\}$ that, each, experience a step-like change in correlation time. In order to build the actual noise protocol driving these trajectories, we generate in parallel two independent sequences of $\eta_t$ time-series with different (but constant) correlation times that are then interspersed synchronously with the $\omega_c^i/\omega_c^f$ square modulation impacting the motional trajectory $x_t$. The detailed procedure is described in Fig. \ref{fig:LongTraj}.

\section{Ergodicity}
\label{APPENDIX_Ergodicity}
\begin{figure}[htb]
	\centering{
			\includegraphics[width=0.35\textwidth]{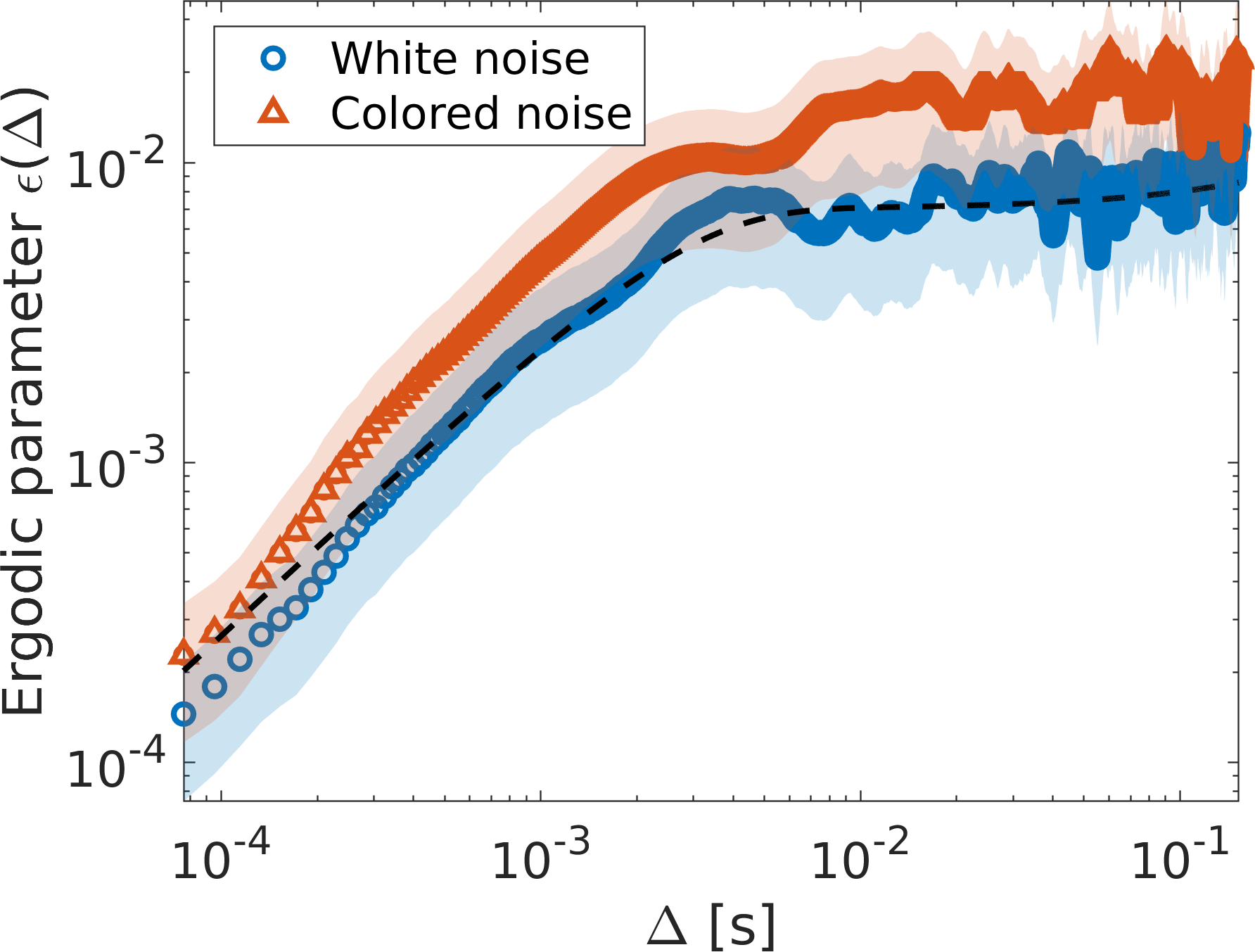}
			\label{fig:Ergodicity}}
		\centering{
			\includegraphics[width=0.35\textwidth]{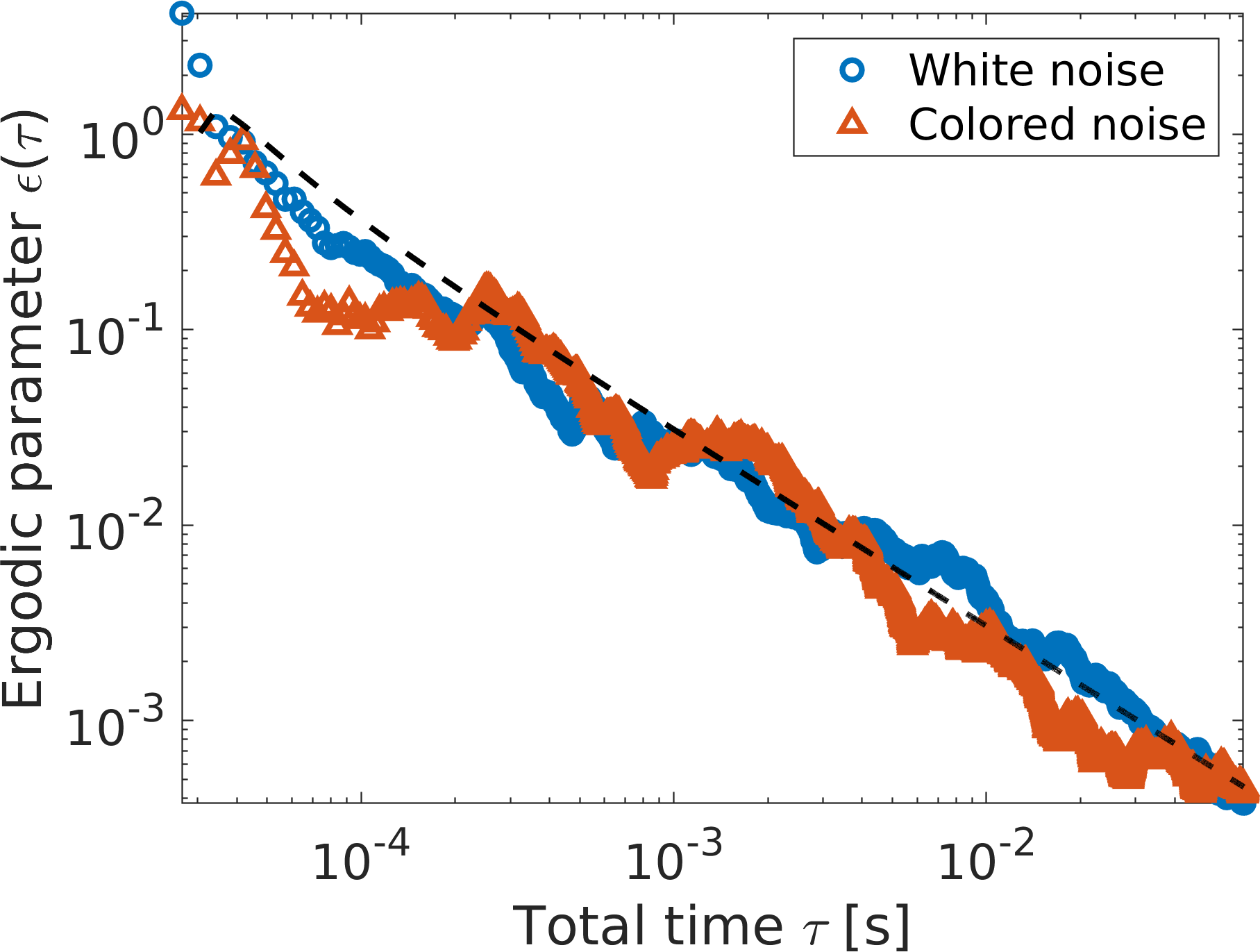}
			\label{fig:ErgodicityVsTime}}
		\caption{{ (a) The ergodic parameter (i.e. the normalized variance $\epsilon(\Delta)$ presented in our previous work \cite{Goerlich2021}) is shown as a function of the time lag $\Delta$ for both the white noise driven process (blue circles) and the colored noise driven process (red triangles) along with the analytical prediction for the white noise case (black dashed line). The red and blue hazes measure the $95\%$ confidence interval.
				(b) The same ergodic parameter $\epsilon(\Delta)$ plotted as a function of the total time $\tau$ for both the white-noise driven process (blue circles) and the colored-noise driven process (red triangles) along with the analytical prediction for the white noise case (black dashed line).
		}}
\end{figure}
\vspace{3mm}

\highlight{For a stationary process, ergodicity is the equality of time and ensemble averages in the limit of infinite time $\mathcal{T}$ and infinite ensemble}. But as detailed in our previous work \cite{Goerlich2021}, appropriate tools exist that can asses the ergodic nature of trajectories $\{x_t\}$ on finite samples and finite times. To do so, we rely on an estimator \cite{METZLER2000} corresponding to the variance of the ratio between time averaged mean squared displacement (MSD) and time-ensemble averaged MSD.
This ratio should becomes Dirac-like for long-time (or short time lag $\Delta$ in the MSD).
The $0$ limit of the variance of this ration for $\Delta/\mathcal{T} \rightarrow 0$ is a necessary and sufficient condition for ergodicity \cite{Goerlich2021}.

The result, plotted in Fig. \ref{fig:Ergodicity} (a) with fixed $\mathcal{T}$ and varying $\Delta$, decays to zero for short time-lag as expected. With fixed $\Delta$, varying $\mathcal{T}$, the expected decrease towards 0 for long time, with a linear trend in log scale, is also clearly seen in Fig. \ref{fig:ErgodicityVsTime} (b).
These two results validate our ergodic assumption for the time-series of position $x_t$, and therefore our treatment when it comes to building trajectory ensembles.

\section{Power spectral density, autocorrelation, mean squared displacement and equipartition breaking}
\label{APPENDIX_MSD}

Our system, consisting of an optically trapped bead thermally diffusing within active fluctuations, is described by a couple of stochastic differential equations that determine the evolution of the position of the bead within the trap according to:
\be
\dot{x}_t = - \omega_0 x_t + \sqrt{2 D}\xi_t + \sqrt{2 D_a}\eta_t
\label{eq:model}
\ee
where the active noise $\eta_t$, solution of the Ornstein-Uhlenbeck process
\be
d\eta_t = -\omega_c \eta_t dt + \sqrt{2 \alpha \omega_c}dW_t,
\label{eq:noise2}
\ee
is an exponentially correlated Gaussian variable.

\begin{figure}[htb]
	\centering{
		\includegraphics[width=0.4\textwidth]{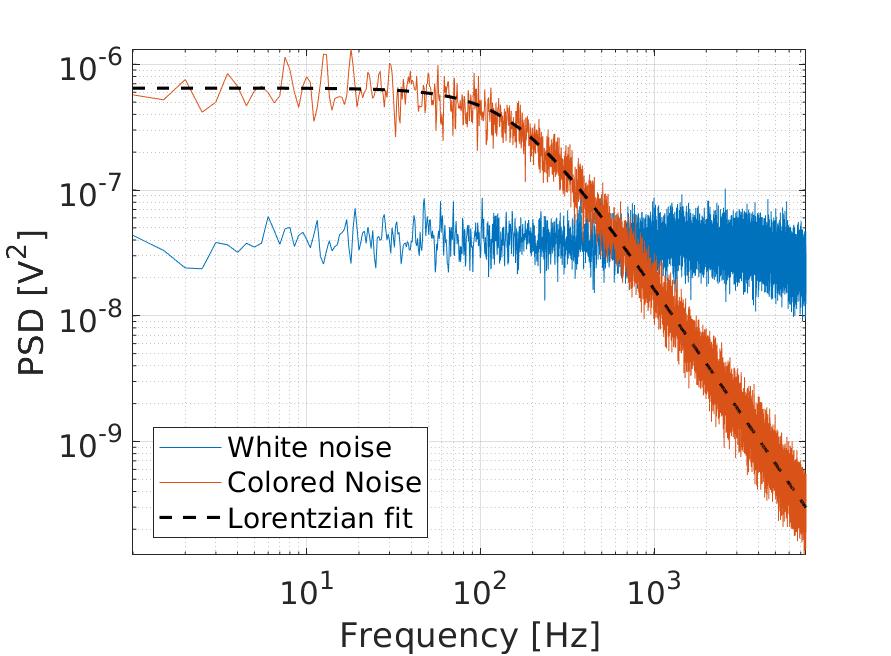}
		\label{fig:PSDinput}}
	\centering{
		\includegraphics[width=0.4\textwidth]{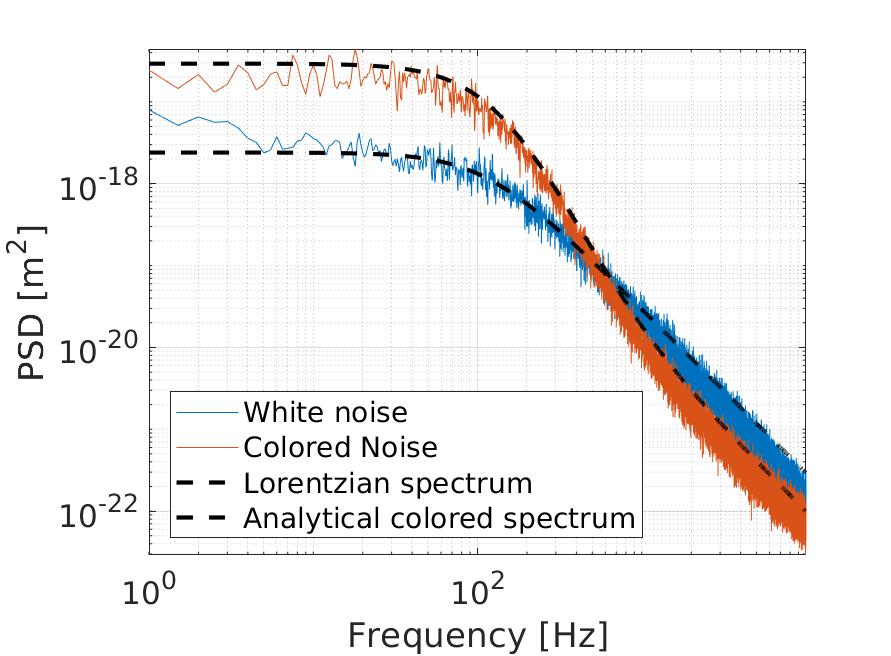}
		\label{fig:PSD}}
	\caption{{ (a) Power spectral densities (PSD) of the white noise (blue curve) and colored noise (red curve) measured on a $10\%$ fraction of the laser beam signal that is sent inside the trap to act on the bead as a noisy radiation pressure. A Lorentzian fit (black dashed curve) is superimposed on the colored noise spectrum.
			(b) Motional PSD $S_x[\omega]=2|x[\omega]|^2$ plotted as a function of the frequency for the white noise driven process (blue) and the colored noise driven process (red) along with the associated theoretical PSD (black dashed curves).
	}}

\end{figure}

We can derive the noise power spectrum density by Fourier transforming Eq. (\ref{eq:noise2})
\be
-i\omega \eta[\omega] = -\omega_c \eta[\omega]  + \sqrt{\alpha \omega_c} \xi[\omega]
\ee
where $\omega_c$ is the correlation pulsation. Taking the squared norm leads to the active noise power spectral density (PSD)
\be
 \eta[\omega] \eta^*[\omega]  =|\eta [\omega]|^2 = \frac{\alpha \omega_c}{\omega_c^2 + \omega^2}.
\label{eq:PSDinput}
\ee
On Fig. \ref{fig:PSDinput} (a), we plot the PSD directly measured from the laser output signal used to induce the noisy radiation pressure, both in the case of a white noise and colored noise.
As expected, the spectrum of the white noise is flat on all the studied bandwidth, whereas the spectrum of the colored noise is following a Lorentzian profile, well captured by a fit following Eq. (\ref{eq:PSDinput}).

\vspace{3mm}

The PSD of the motion $x_t$ is evaluated from Eq. (\ref{eq:model}) as:
\be
 {x}[\omega]{x}^*[\omega]  = \frac{1}{\omega_0^2 + \omega^2} \left( 2D {\xi}[\omega]{\xi}^*[\omega]  + 2D_a {\eta}[\omega]{\eta}^*[\omega]   \right),
\label{eq:process1}
\ee
noting that the implicit averaging performed in this square cancels the two cross-product of the uncorrelated noises $\eta[\omega] \xi ^*[\omega] $ and the complex conjugate. In Eq. (\ref{eq:process1}), $\omega_0 = \kappa/\gamma$ is the inverse of the characteristic relaxation time of the system, and $ \eta[\omega]\eta^*[\omega] $ is given by Eq. (\ref{eq:PSDinput}) and $ \xi[\omega]\xi^*[\omega] = 1$. Hence
\be
{x}[\omega] {x}^*[\omega]  = | x[\omega]|^2= \frac{1}{\omega_0^2 + \omega^2} \left( 2D + \frac{2D_a\alpha\omega_c}{\omega_c^2 + \omega^2} \right)
\label{eq:process2}
\ee
On Fig. \ref{fig:PSD} (b), we plot the measured spectra of $x_t$ both for a white and colored external drive, with the analytical result of Eq. (\ref{eq:process2}) using the value of $D_a$ obtained from the fit of the MSD for the colored noise (see Fig. \ref{fig:SomeMSDWithoutFit} above) and using the $D_a \rightarrow 0$ limit for the white noise case.
A very good agreement between the theory and the experimental data is clearly seen, confirming that our model captures well the real diffusive dynamics of the trapped bead.

We can also compute the correlation function of the colored noise driven process from the Wiener-Khintchine theorem as:
\be
\begin{aligned}
	C_{xx}(\Delta) & = \frac{1}{2 \pi} \int_{-\infty}^{+\infty} | x[\omega]|^2 e^{-i\omega\Delta} d\omega\\
	& = \frac{1}{2 \pi} \int_{-\infty}^{+\infty} \frac{2 D e^{-i\omega\Delta} d\omega}{\omega_0^2 + \omega^2}  +  \frac{1}{2 \pi} \int_{-\infty}^{+\infty} \frac{2 D_a \alpha \omega_c e^{-i\omega\Delta} d\omega}{\left(\omega_0^2 + \omega^2\right)\left( \omega_c^2 + \omega^2\right)},
\end{aligned}
\ee
where both integrals can be computed via contour integration. For the first one, $f[\omega] = \frac{D}{\pi} \frac{e^{-i\omega\Delta}}{\omega_0^2 + \omega^2}$ has one simple pole in the upper-half complex plane in $i\omega_0$, leading to compute one residue
\be
\begin{aligned}
	\int_{-\infty}^{+\infty} f[\omega] d\omega &= 2 i \pi Res\{ f[\omega], i\omega_0 \}\\
	&= \lim_{\omega\to\omega_0} \frac{2 D e^{-i\omega\Delta}}{\omega + i\omega_0}\\
	&= \frac{D}{\omega_0} e^{-\omega_0\Delta}.
\end{aligned}
\ee
Similarly, the second integral with $g[\omega] = \frac{D_a \alpha \omega_c e^{-i\omega\Delta} }{\pi \left(\omega_0^2 + \omega^2\right)\left( \omega_c^2 + \omega^2\right)} $ is evaluated by separating it through partial fraction decomposition in $g[\omega] = \frac{D_a \alpha \omega_c e^{-i\omega\Delta}}{\pi(\omega_c^2 - \omega_0^2)} \left( \frac{1}{\omega_0^2 + \omega^2} - \frac{1}{\omega_c^2 + \omega^2} \right) \equiv g_1[\omega] + g_2[\omega]$
leading to two integrals with simple poles in $i\omega_0$ and $i\omega_c$
\be
\begin{aligned}
	\int_{-\infty}^{+\infty} g[\omega]d\omega &= 2 i \pi Res\{ g_1[\omega], i\omega_0 \} + 2 i\pi Res\{ g_1[\omega], i\omega_c\}\\
	&= D_a\alpha\omega_c^2 \left( \frac{e^{-\omega_0\Delta}}{\omega_0\left(\omega_c^2 -\omega_0^2\right)}  - \frac{e^{-\omega_c\Delta}}{\omega_c\left(\omega_c^2 -\omega_0^2\right)} \right)\\
	&= \frac{D_a \alpha \omega_c}{\omega_0 (\omega_c^2 - \omega_0^2)}\left(e^{-\omega_0 \Delta} - \frac{\omega_0}{\omega_c} e^{-\omega_c\Delta}\right).
\end{aligned}
\ee
These evaluations are combined to provide the expression for the correlation function of the diffusion process:
\be
C_{xx}(\Delta) = \frac{D}{\omega_0} e^{-\omega_0\Delta} + \frac{D_a \alpha \omega_c}{\omega_0 (\omega_c^2 - \omega_0^2)}\left(e^{-\omega_0 \Delta} - \frac{\omega_0}{\omega_c} e^{-\omega_c\Delta}\right)
\ee

\begin{figure}[htb]
	\centering{
		\includegraphics[width=0.6\linewidth]{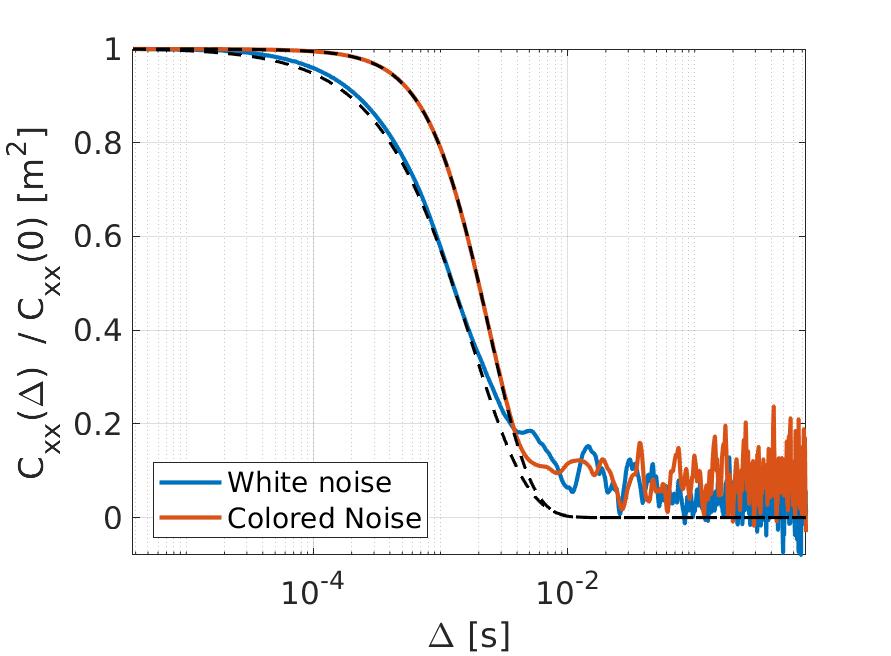}}
	\caption{{ Correlation function $C_{xx}(\Delta) = \langle x(t + \Delta) x(t)\rangle$ plotted as a function of the lag $\Delta$ for the white (blue curve) and colored (red curve) noise-driven processes, both normalized to the zero-delay $\Delta = 0$ correlation function $C_{xx}(0)$. The theoretical expressions derived in Sec. \ref{APPENDIX_MSD} are displayed as dashed black curves for both cases.
	}}
	\label{fig:correlation}
\end{figure}
\vspace{3mm}

On Fig. \ref{fig:correlation} we represent the normalized correlation function $C_{xx}$ for both white and colored noise drives where we superimpose the analytical result, using the value of $D_a$ obtained, as indicated above, from the fit of the MSD for the colored noise. Here too, we use the $D_a \rightarrow 0$ limit for the white noise. Again, the good agreement between the exponential decays and the analytical models is observed.

The MSD of a colloid diffusing in a thermal environment obeys an Ornstein-Uhlenbeck process and is thus characterized by the white noise MSD:
\be
\langle \delta x^2(\Delta) \rangle \equiv \langle (x(t+\Delta) - x(t))^2 \rangle = \frac{2D}{\omega_0} \left( 1 - e^{-\omega_0 \Delta}\right)
\label{eq:MSD_whitenoise}
\ee
where $D$ is the diffusion coefficient in the thermal bath, expressed in $\si{\meter^2/\second}$, and $\omega_0 = \kappa/\gamma$ is the inverse of the characteristic relaxation time of the bead in the trap.

In contrast, the MSD of an active particle obeys Eq. (\ref{eq:model}) and can be computed as $\langle \delta x^2(\Delta) \rangle = 2 \langle x_t^2 \rangle - 2 C_{xx}(\Delta)$ where the variance is the stationary variance of the process and $\langle x_t^2 \rangle$ is taken as the limit $\lim\limits_{t \rightarrow \infty} [\langle x_t^2 \rangle]= \frac{D}{\omega_0} + \frac{D_a \alpha}{\omega_0(\omega_c + \omega_0)}$. This leads us to calculate directly:

\be
\begin{aligned}
	\langle \delta x^2(\Delta) \rangle =& \frac{2D}{\omega_0} + \frac{2D_a \alpha}{\omega_0(\omega_c + \omega_0)} - \frac{2D}{\omega_0} e^{-\omega_0\Delta} - \frac{2D_a \alpha \omega_c}{\omega_0 (\omega_c^2 - \omega_0^2)}\left(e^{-\omega_0 \Delta} - \frac{\omega_0}{\omega_c} e^{-\omega_c\Delta}\right)\\
	&= \frac{2D}{\omega_0} \left(1 - e^{-\omega_0 \Delta}\right) + \frac{2 D_a \alpha \omega_c}{\omega_0(\omega_c^2 - \omega_0^2)} \left(1 - e^{-\omega_0 \Delta} - \frac{\omega_0}{\omega_c} (1 - e^{-\omega_c \Delta}) \right)
\end{aligned}
\label{eq:MSD_colorednoise}
\ee
that constitutes the result used in the main text.

The long-time limit can be easily derived as:
\be
\lim\limits_{\Delta \rightarrow \infty}[\langle \delta x^2(\Delta) \rangle] = \frac{2D}{\omega_0} + \frac{2 D_a\alpha}{\omega_0(\omega_c + \omega_0)}.
\label{eq:MSD_limit}
\ee

To asses the break of the equipartition relation, it is enough to show the absence of linearity between the variance and the effective diffusion coefficient associated with the active process.
For different noise colors, we thus measure and fit the MSD to extract each corresponding $D_a$. We then plot in Fig. 2, panel (b) of the main text the variance against the total diffusion coefficient $D + D_a$. As seen in the figure, we clearly observe that the variance does not follow the intuitive $\omega_0^{-1}$ linearity. In striking contrast, it rather follows one half of Eq. (\ref{eq:MSD_limit}) -- i.e., $\frac{D}{\omega_0} + \frac{ D_a\alpha}{\omega_0(\omega_c + \omega_0)}$ -- where the explicit $\omega_c$ term prevents us from defining a unique effective diffusion coefficient (or effective temperature) and where $D_a$ comes from the fit of the MSD displayed in Fig. \ref{fig:SomeMSDWithoutFit}.

\section{Micro Rheology and Fluctuation Dissipation Theorem}
\label{APPENDIX:FDT}

We probe the non-equilibrium nature of the active fluctuations and the validity of the Fluctuation Dissipation Theorem (FDT) by comparing the dynamical responses of our system to Active MicroRheological (AMR) and Passive MicroRheological (PMR) excitations, respectively \cite{Turlier2016, Chakrabarti2019}.

At thermal equilibrium, under detailed balance conditions, the linear response of the system to a small perturbation is connected to equilibrium correlations of fluctuations through the FDT according to which:
\be
\frac{\partial C_{xx}(t)}{\partial t} = 2 k_B T R(t),
\ee
where $C_{xx}(t) = \langle x(t) x(0) \rangle$ is the motional autocorrelation function and $R(t)$ is the response function of the system.
This equation can be more conveniently derived in the frequency domain.
If we consider the motion of the bead driven by a noise of unit variance $\phi_t$ (that takes in our case the form $\xi_t + \sqrt{D_a/D}\eta_t$), the Fourier transform of the corresponding Langevin equation writes as:
\be
-i\omega \gamma {x}[\omega] =  -\kappa {x}[\omega] + \sqrt{2k_BT\gamma}\phi [\omega],
\ee
where $\kappa$ is the stiffness of the potential, $\gamma$ the Stokes friction drag, and $\sqrt{2k_BT\gamma}\phi [\omega]$ is a generic random force.
The equation can be written in terms of a mechanical susceptibility $\chi [\omega]$ as
\be
{x}[\omega] = \chi[\omega] \sqrt{2k_BT\gamma}\phi [\omega]
\ee
where $\chi [\omega]$ can be decomposed into real and imaginary parts as:
\be
\chi[\omega] = \frac{\omega_0}{\gamma(\omega_0^2 + \omega^2)} + i \frac{\omega}{\gamma(\omega_0^2 + \omega^2)} \equiv \chi'[\omega] + i \chi''[\omega].
\ee
If we compare the imaginary part $\chi''[\omega]$ with the power spectral density obtained, in the case of a thermal, white noise drive, by the square modulus of position Fourier transform $|x[\omega]|^2 = 2 D /(\omega_0^2 + \omega^2)$, we obtain the expression of the FDT in the Fourier space:
\be
\chi''[\omega] = \frac{\omega |x[\omega]|^2}{2 k_B T}  \label{eq:chiPSD}
\ee
where the spectrum $|x[\omega]|^2$ is the Fourier transform of autororrelation function $C_{xx}(t)$ (Wiener-Khinchine theorem).

If now one adds a small sinusoidal perturbation on the bead by means of an external force (which  corresponds to radiation pressure in our experiments), the FDT can be tested experimentally  by measuring the response function. Under the sinusoidal \textit{ac} drive of the AMR mode at pulsation $\omega_{ac}$, the PSD takes the following form \cite{Schnoering2019}
\be
|x[\omega]|^2_{ac} = \frac{1}{\omega_0^2 + \omega^2}\left( 2D + \frac{F_{ac}^2}{2\gamma^2}\delta(\omega - \omega_{ac}) \right)
\ee
where $F_{ac}$ is the Fourier force component of the drive, while the unperturbed PSD of the PMR mode writes
\be
|x[\omega]|^2= \frac{2D}{\omega_0^2 + \omega^2}.
\ee

By computing the ratio $|x[\omega_{ac} ]|^2/|x[\omega_{ac} ]|^2_{ac} $ at the pulsation $\omega_{ac}$ for a bead on which a ``white noise'' radiation pressure is exerted (hence maintaining it close to thermal equilibrium but at an effective temperature $T_{\rm eq}$ higher than room temperature), we can extract the value of $F_{ac}$ by taking the mean value of all realizations. This value can then be used to calibrate the response function $\chi'' [\omega]$ and compare it with the steady-state fluctuation PSD $2|x[\omega]|^2$.

\begin{figure}[htb!]
	\centering{
			\includegraphics[width=0.4\textwidth]{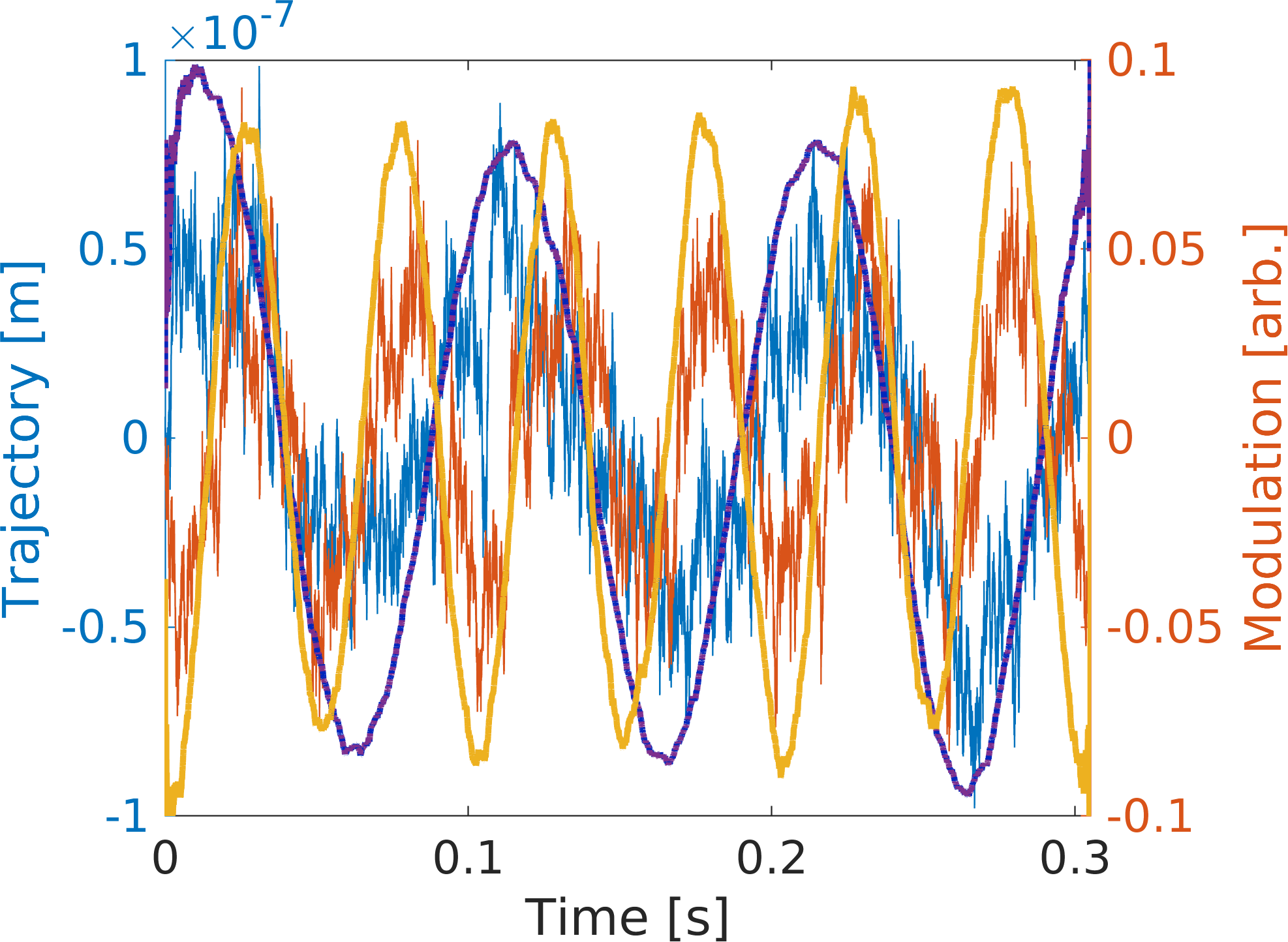}
			\label{fig:Microrheo}}
		\centering{
			\includegraphics[width=0.4\textwidth]{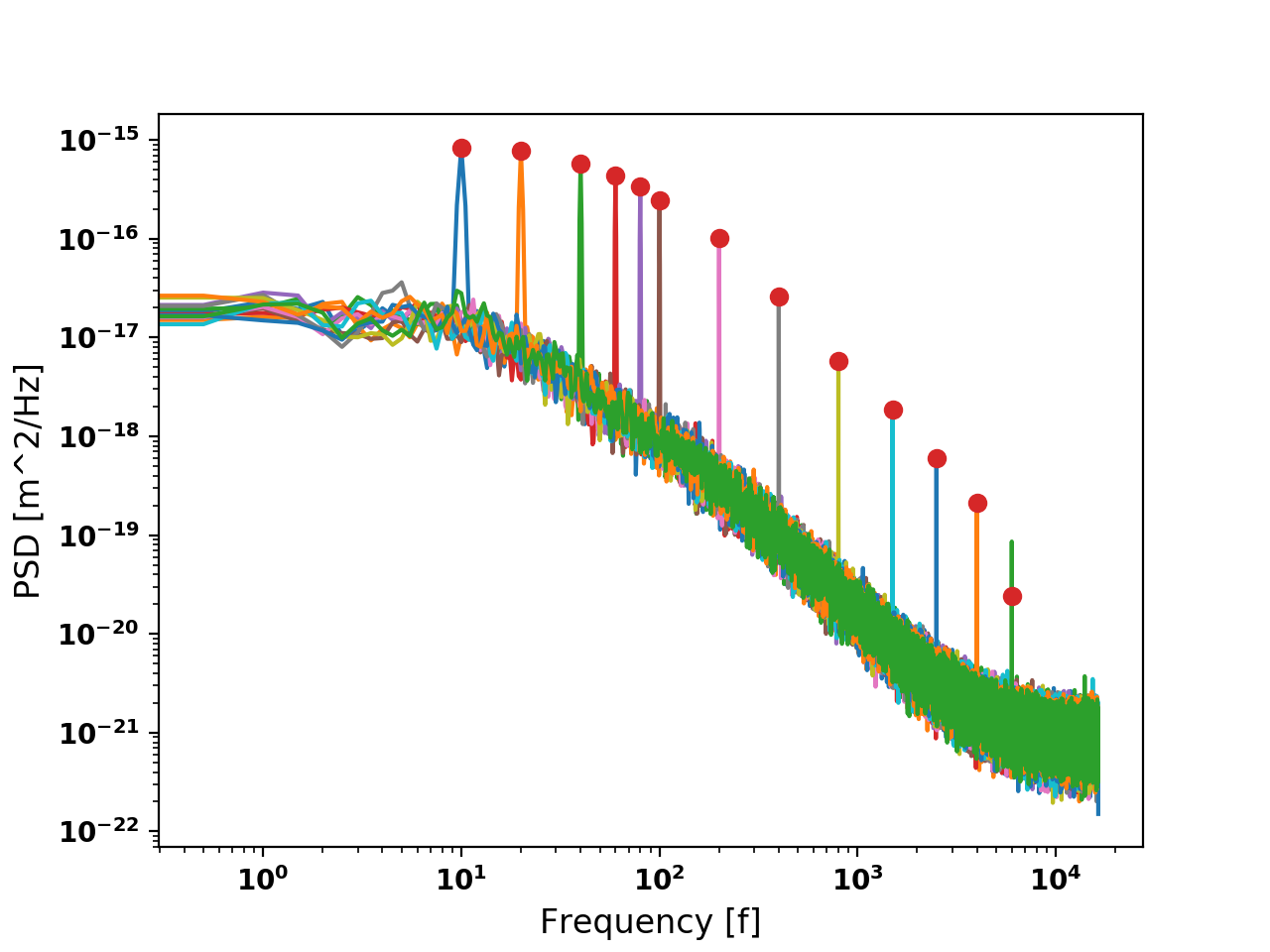}
			\label{fig:CnSinePSD}}
		\caption{{ (a) Active micro-rheological (AMR) experiment where the sinusoidal forcing of the system is monitored in the time domain. The recorded trajectories of the bead inside the trap are superimposed to the sinusoidal traces of the force for two different modulation frequencies.
				(b) Power spectral densities displayed together for different modulation frequencies of the external force drive. The Fourier components of each harmonic forcing are clearly seen as peaks in the PSD.
		}}
\end{figure}
\vspace{3mm}

\begin{figure}[htb!]
	\centering{
		\includegraphics[width=0.5\linewidth]{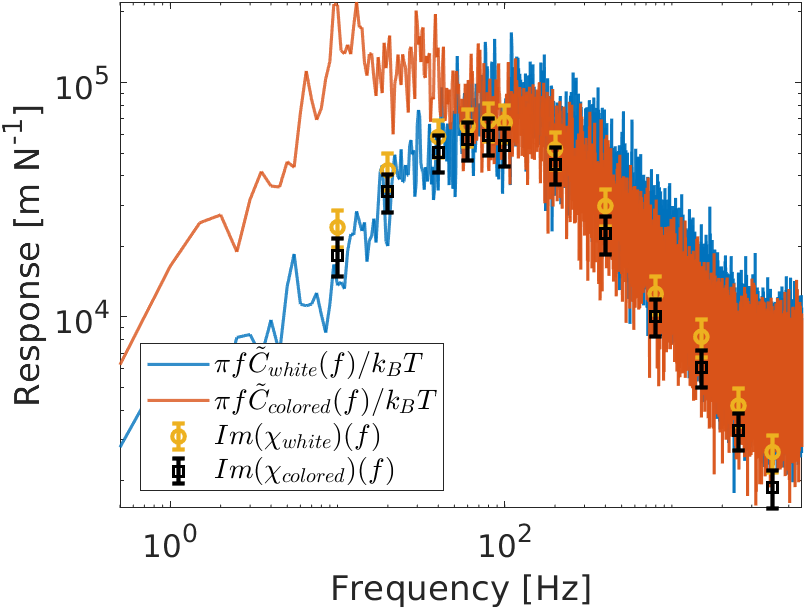}}
	\caption{{We compare the measured values of $\chi '' [\omega]$ for white (open circles) and correlated (open squares) noise for different modulation frequencies $\omega_{ac}$ and small sinusoidal perturbations with the stationary correlation spectra plotted as $\omega |x[\omega]|^2 / 2k_B T$ for white (blue curve) and correlated (orange curve) noises.  One immediately remarks the breaking of the FDT for the colored-noise driven process.
	}}
	\label{fig:FDT}
\end{figure}

On Fig. \ref{fig:Microrheo} (a), we show the external drive in the time-domain and the motional response of the bead inside the trap for two different modulation frequencies. By repeating the procedure for frequencies ranging from $10$ Hz to $6$ kHz, the response of the bead is characterized over all the useful bandwidth (see Fig. \ref{fig:CnSinePSD} (b)). On Fig. \ref{fig:FDT}, the values of $\omega |x[\omega]|^2 / 2k_B T$ and $\chi'' [\omega]$ are plotted together for the probed frequencies, for both the white noise and colored noise driven processes. We clearly observe that in both cases, the response functions associated with the mechanical susceptibilities fall back on the same trend. This trend is exaclty the one associated with the white noise driven PSD as expected from Eq. (\ref{eq:chiPSD}). In contrast, the spectral density of the colored noise driven process significantly departs from the FDT in  Eq. (\ref{eq:chiPSD}), and more particularly for the low frequencies of the active fluctuation spectrum.
This is in agreement with other observations made recently in active systems \cite{Turlier2016}, where the active mechanical processes mostly appear at low frequency, while the FDT is recovered for the thermally dominated high-frequency part.

\vspace{3mm}

We note here that a simple observation of the breaking of the FDT can already be seen in our model described by Eq. (\ref{eq:model}), where the fluctuating forces associated with the added noises $\xi_t$ and $\eta_t$ now possess an intrinsic correlation time due to the correlated nature of $\eta_t$, while the friction kernel $\gamma$ is taken as instantaneous $\Gamma(t, t') = \gamma \delta(t,t')$.
This choice has been shown to be valid in the experimental case \cite{Maggi2017}, where  the fluctuations of the active bath are not compensated by a dissipation with the same rate.
In the limit of vanishing correlation times, the FDT is recovered as the noise is white ($\delta$-correlated) and its only effect is an effective change in temperature, as was already observed  using a different experimental technique \cite{Petrov2013, Roldan2014}.

\highlight{
\section{Variance under step-like change of color}
}
\label{APPENDIX:VarStep}

We carefully analyze here the \textit{color protocol} associated with a sudden change (STEP) in the correlation time of the bath.
We model this STEP protocol by applying two different noises $\eta_1$ and $\eta_2$ with respective inverse correlation times $\omega_1$ and $\omega_2$ and modulated with two Heaviside functions.
We consider that at time $t=0^-$ the system is at thermal equilibrium with no additional noise ; at time $t=0^+$, a first colored noise $\eta_1$ is turned on ; at time $t = t_0$, $\eta_1$ is turned off and the second noise $\eta_2$ is turned on.
The differential equation for the centered process $x_t$ writes as:
\be
	\dot{x}_t = - \omega_0 x + \theta(t_0 - t) \sqrt{2 D_1} \eta_1(t) + \theta(t - t_0) \sqrt{2 D_2} \eta_2(t) + \sqrt{2D} \xi(t) ,
\ee
where $\theta(t_0 - t)$ is a Heaviside function centered at $t_0$ and $\xi$ is the thermal white noise. Laplace transforms lead to compute the solution for the position $x_t$ of the microsphere that experiences a STEP-like change of correlation time
\be
	x_t = x_0 e^{-\omega_0 t} + e^{-\omega_0 t}  \left[ \sqrt{2D_2} \int_{t_0}^t e^{\omega_0\tau} \eta_2(\tau)d\tau + \sqrt{2D_1} \int_{0}^{min(t_0 , t)} e^{\omega_0\tau} \eta_1(\tau)d\tau +  \sqrt{2D} \int_0^t e^{\omega_0 \tau} \xi(\tau)d\tau\right].
	\label{eq:StepSolution}
\ee
The first noise $\eta_1$ stops at $t_0$ so that the integral stops either at $t$ if $t<t_0$ or at $t_0$ otherwise.
The second noise $\eta_2$ starts at $t_0$ and so does its integral.
The thermal noise is always present, hence its integral extends from $0$ to $t$.
Note that even if the integral of the first noise stops at $t_0$, we will see on the variance that it still has an influence on longer times.

\begin{figure}[htb]
	\centering{
		\includegraphics[width=0.6\linewidth]{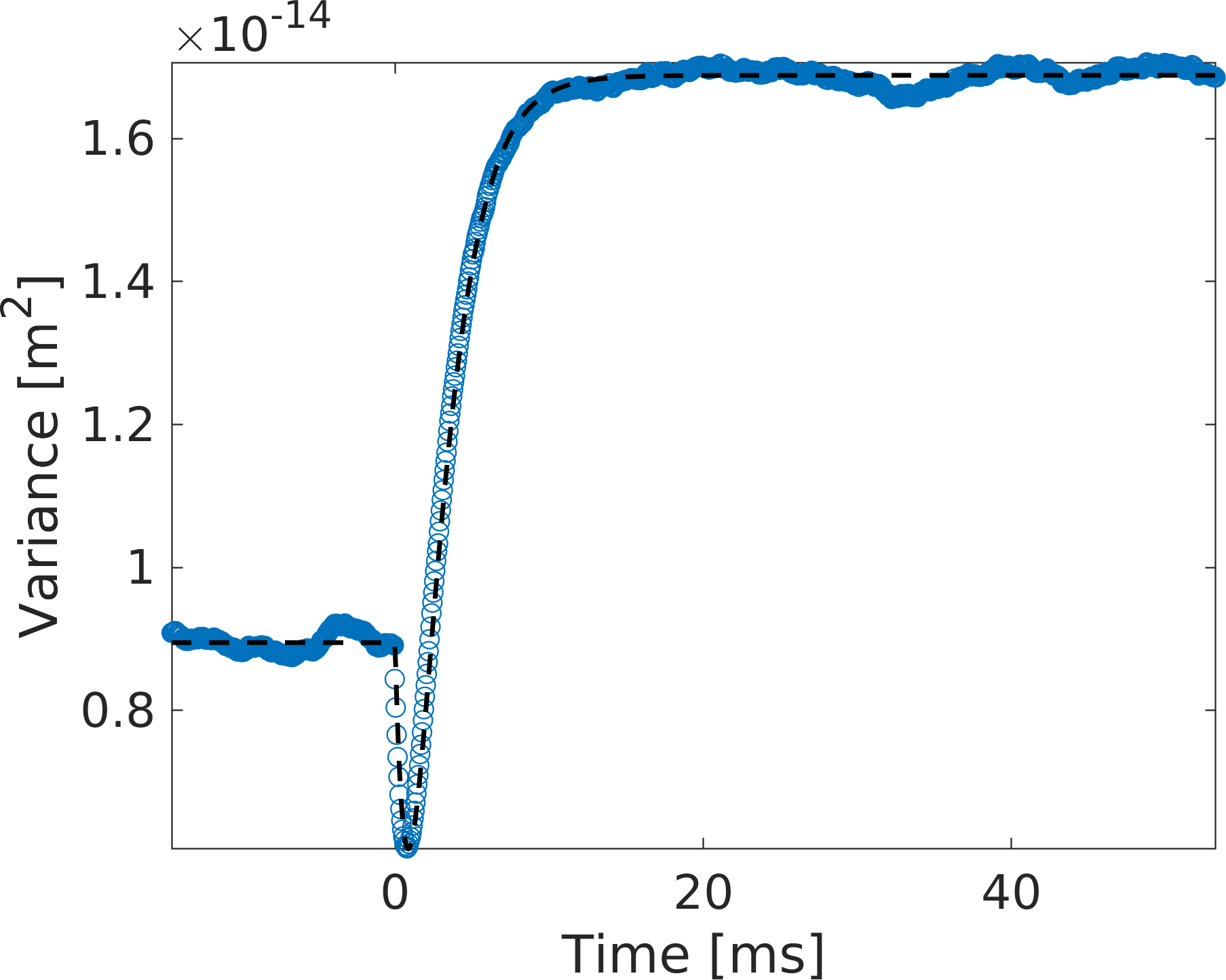}}
	\caption{{Analytical solution for the variance associated with a STEP-like change of color (from $\omega_1^{-1} = 2 ~\si{\milli \second}$ and $\omega_2^{-1} = 40~\si{\milli \second}$, with $\omega_0^{-1} = 2.1~\si{\milli \second}$, here $D_1 = D_2 = 2\times 10^{-9}~ \si{\square\meter/\second}$) along with a numerical simulation performed with a first order Euler-Maruyama discretization scheme.
	}}
	\label{fig:VarFitSIM}
\end{figure}
\vspace{3mm}

We can compute the motional variance as the ensemble average of the square of Eq. \ref{eq:StepSolution}.
By a parametrization of time with $\Delta$ corresponding to the lag after the STEP, we arrive for the time-evolution of the variance after the protocol at:
\be
	\sigma^2_x(\Delta) = \frac{D}{\omega_0} + \frac{D_1 \alpha }{\omega_0(\omega_0 + \omega_1)} e^{-2\omega_0 \Delta} + \frac{2 D_2 \alpha}{\omega_0^2 - \omega_2^2} \left[1 - 2 e^{- \Delta (\omega_0 + \omega_2)} + e^{-2\omega_0 \Delta} - \frac{\omega_2}{\omega_0}(1 - 2e^{-2\omega_0 \Delta})\right] ,
\label{Eq:VarAfter}
\ee
while we recall the stationary variance before the STEP when $t_0$ is very large
\be
	\sigma^2_x = \frac{D}{\omega_0} + \frac{D_1 \alpha}{\omega_0(\omega_0 + \omega_1)}.
\label{Eq:VarBefore}
\ee
We can note on Eq. \ref{Eq:VarAfter} that the influence of the first noise $\eta_1$ decreases exponentially with the characteristic time $1/\omega_0$ while the second noise emerges with a non-monotonic bi-exponential relaxation.
On Fig. \ref{fig:VarFitSIM} we show the theoretical variance as the combination of Eq. \ref{Eq:VarBefore} and \ref{Eq:VarAfter} with the result of a numerical simulation with the same parameters.
We see that the model captures well the dynamics at play, including the decrease of variance after the protocol, due to the short-time underestimation of the correlation (mathematically, to the combination of the exponential terms in the variance solution).

\section{Thermodynamics}
\label{APPENDIX:Thermo}

The application of stochastic thermodynamics to active matter has already been studied theoretically \cite{Chakrabarti2018, Chakrabarti2019, Speck2016, Marconi2017, Eichhorn2019} and experimentally in some cases \cite{Volpe2016}. In this Appendix, we describe in detail how the stochastic heat can be efficiently used to describe and characterize the processes at play in our experiments.
We first note that, in our experiments, our system is brought to a Non-Equilibrium Steady-State (NESS) where the stationary stochastic laser drive maintains -- through the action of radiation pressure -- the system out of its equilibrium state at a given temperature and stiffness $\kappa$.
Following the standard methods of stochastic energetics \cite{Sekimoto1998, SekimotoBook}, we write, from Eq. (\ref{eq:model}), for our process

\be
	\left( \gamma \dot{x}_t - \gamma\sqrt{2 D}\xi_t dt \right) dx = - \left(\kappa x_t + \gamma \sqrt{2 D_a}\eta_t\right) dx.
	\label{eq:ModelThermo}
\ee
The left-hand side is interpreted as the heat exchanged with the thermal bath $\delta q = -\left( \gamma \dot{x} - \sqrt{2 D}\xi_t dt \right) dx $.
Since the active force is stochastic, it produces no work. In fact, including a random force in the expression of work induces a violation of the Crooks relation \cite{Gomez_Solano_2010}. The internal energy stays related to the potential energy $dU = - \kappa x_t^2 dx$ and the remaining term $\gamma \sqrt{2 D_a}\eta_t dx$ is the energy exchanged with the active bath. Interpreted as a heat term \cite{Eichhorn2019}, it can be evaluated from the right-hand side of Eq. (\ref{eq:ModelThermo}) as
\be
	\delta q(t) = \frac{1}{2}\kappa \frac{d  x^2 }{dt}dt - \gamma\sqrt{2 D_a}   \eta_t \dot{x}_t dt ,
\ee
which can be integrated to give the stochastic heat evaluation
\be
	q(t) = \frac{1}{2}\int_0^t\kappa \frac{dx_s^2 }{ds}ds - \gamma \int_0^t \sqrt{2 D_a} \eta_s \dot{x_s} ds.
\ee

Finally, we compute the ensemble average heat, which will be expressed in terms of variance and cross-correlations
\be
	Q(t) \equiv \langle q(t) \rangle = \frac{1}{2}\int_0^t\kappa \frac{d \langle x_s^2 \rangle}{ds} ds - \gamma \int_0^t\sqrt{2 D_a} \langle \eta_s \dot{x_s}\rangle ds.
\ee
The first term is connected to the evolution of the variance. It vanishes in the steady-state and only accounts for the heat released during a transient evolution of the distribution.
The second term can be computed analytically by injecting the differential equation Eq. (\ref{eq:model}) for $\dot{x}_t$ : $\langle \dot{x_t} \eta_t \rangle = -\omega_0\langle x_t \eta_t \rangle +  \sqrt{2 D_a} \alpha$.
The first term can be computed
\be
	\begin{aligned}
		\langle x_t \eta_t \rangle &=  \int_0^t\sqrt{2 D_a} \langle \eta_t \eta_s\rangle e^{-\omega_0(t - s)}ds\\
		&= \alpha \int_0^t\sqrt{2 D_a} e^{-\omega_c |t - s|-\omega_0(t - s)}ds\\
		&= \frac{\sqrt{2D_a} \alpha}{\omega_0 + \omega_c} \left( 1 - e^{-t(\omega_0 + \omega_c)} \right),
	\end{aligned}
\ee
and the second term
\be
	\begin{aligned}
		-\gamma \int_0^t\sqrt{2D_a}\langle \dot{x}_s\eta_s \rangle ds &= -\gamma \int_0^t\sqrt{2D_a} \left(-\omega_0 \langle x_s\eta_s \rangle + \sqrt{2D_a}\alpha\right)ds\\
		&= 2 \gamma D_a \alpha \left(\int_0^t \frac{\omega_0} {\omega_0 + \omega_c} \left[1 - e^{-s(\omega_0 + \omega_c)}\right]ds - t\right)\\
		&= 2 \gamma D_a \alpha \left(\frac{\omega_0 }{\omega_0 + \omega_c} -1\right) t + \frac{2 \gamma D_a \omega_0}{(\omega_0 + \omega_c)^2}(1 - e^{-t(\omega_0 + \omega_c)}),
	\end{aligned}
\ee
to give, after an exponential decorrelation at short times (just after the noise is turned on, a short-time regime that is never probed in our experiments), a linear heat expenditure with negative (since $\omega_0 > 0$) slope $2 \gamma D_a \alpha \left(\frac{\omega_0 }{\omega_0 + \omega_c} -1\right)$ that account for the heat needed to maintain the system in its NESS.
\highlight{
We note again that, in the white noise limit on an infinite bandwidth, $\omega_c \rightarrow \infty$, if we keep in mind that $D_a \sim \omega_c^{-1}$, housekeeping heat vanishes.
Keeping finite $\omega_c$ we have the expected trend of a housekeeping that increases with correlation time.
}

Therefore, if we discard the decorrelation after the noise is turned on, we obtain the following expression for the cumulative heat:
\be
	\begin{aligned}
		Q(t) &=  \frac{1}{2}\int_0^t\kappa \frac{d \langle x_{s}^2 \rangle}{ds} ds + 2 \gamma D_a \alpha \left(\frac{\omega_0 }{\omega_0 + \omega_c} -1\right) t\\
		&\equiv Q_{EX}(t) + Q_{HK}(t),
	\end{aligned}
\ee
where the two quantities are reminiscent of the \textit{excess} (EX) and \textit{housekeeping} (HK) heat terms \cite{Seifert2012}.

 \begin{figure}[htb]
	\centering{
			\includegraphics[width=0.4\textwidth]{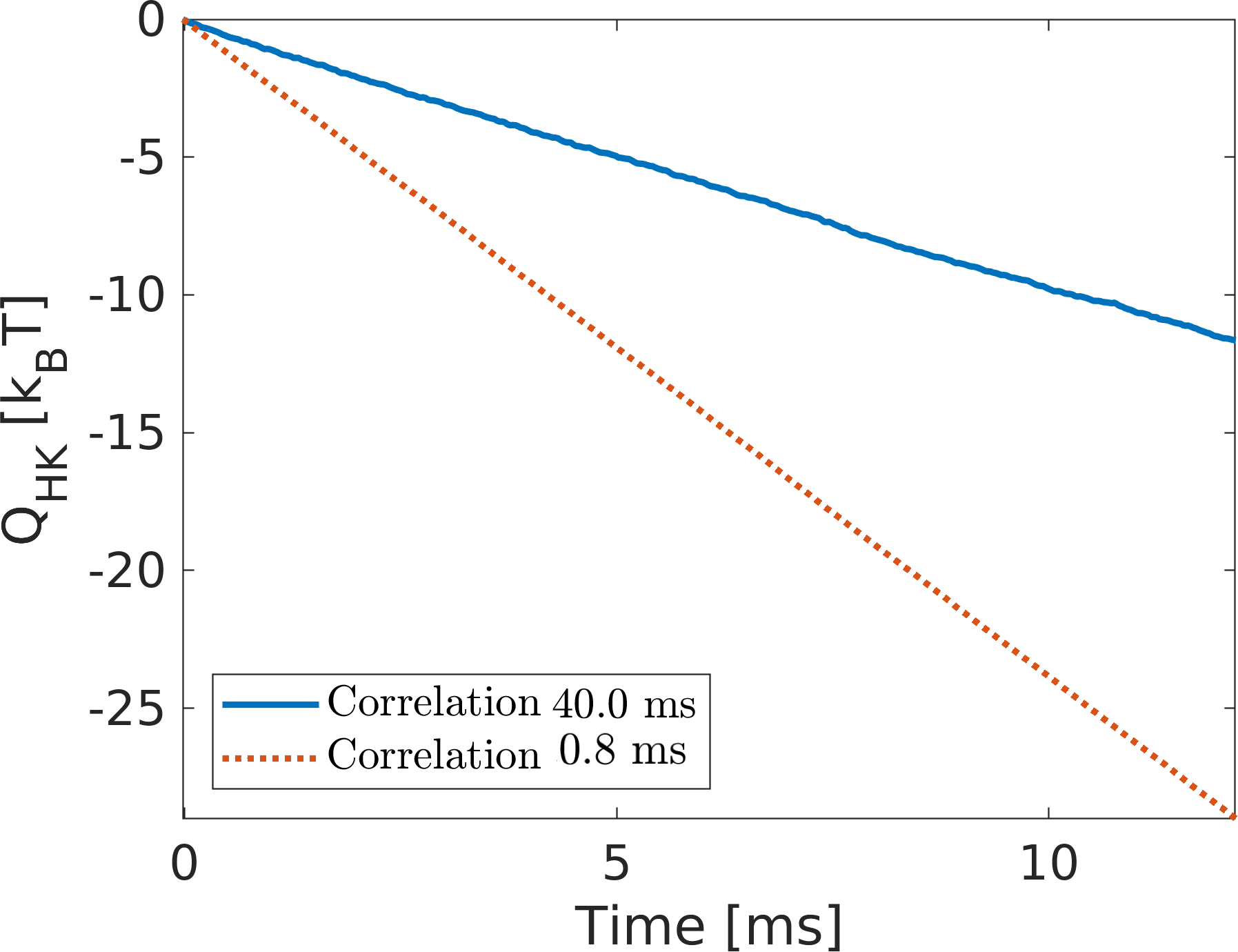}
			\label{fig:Heat2}}
		\centering{
			\includegraphics[width=0.4\textwidth]{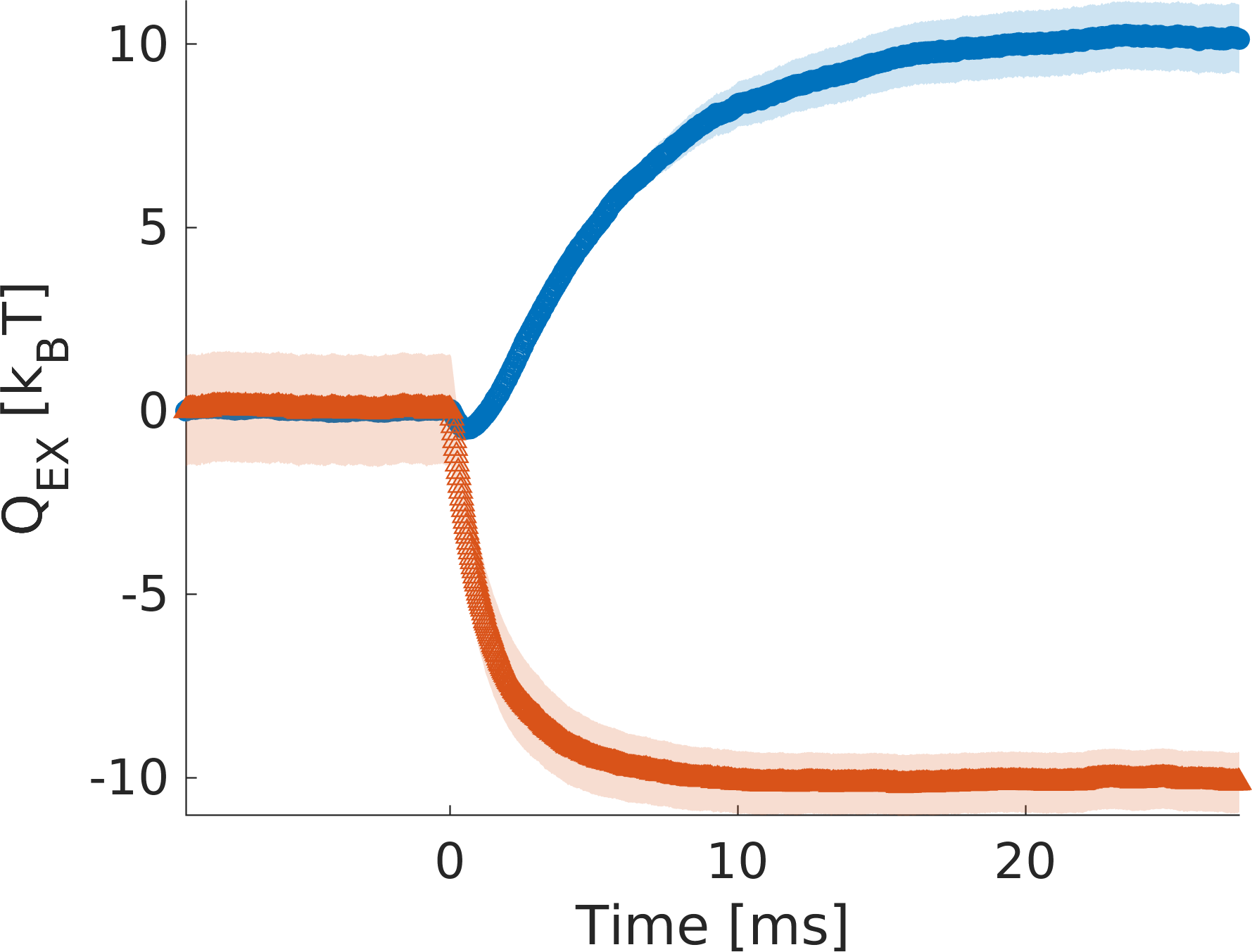}
			\label{fig:Heat1}}
		\caption{{ (a) Measured heat necessary to keep the system in a NESS (in units of $k_B T$) both for $\tau_c = 0.8$ ms (blue line, before the STEP) and for $\tau_c = 40$ ms (red dotted line, after the STEP).
				(b) Released heat measured through the transient both for a increasing correlation time (blue circles) and equivalently for a decreasing correlation time (red triangles).
		}}
	\label{fig:Heat}
\end{figure}
\vspace{3mm}

In Fig. \ref{fig:Heat}, we display the time evolution of these two quantities for the $\tau_c(t)$ STEP protocol described above. Figure \ref{fig:Heat2} (a) shows the heat necessary to maintain the NESS both before and after the change of $\tau_c$. As we see, changing the correlation time changes the rate of heat dissipation. In Fig. \ref{fig:Heat1} (b), the time-evolution of the excess heat discarded through the transient is plotted for both increasing or decreasing STEP of $\tau_c$. It is remarkable to stress that the quantity of heat $\approx 10 k_B T$ exchanged is the same for both cases.

Note that an alternative expression for the heat can be found in the context of active matter and active Ornstein-Uhlenbeck processes.
These different approaches lead all to similar results, since Sekimoto's definition of heat as
\be
\delta q = -\left( \gamma \dot{x} - \gamma\sqrt{2 D}\xi_t dt \right) dx
\ee
is uniquely defined \cite{Sekimoto1998,Park2020}. The differences stem from the way to evaluate this quantity.
In the description of non-reciprocal systems \cite{Loos2020}, the steady-state heat is computed as a sum of correlation between variables and velocities. In our case of unidirectional coupling, this simplifies to a term $\sim \langle \eta_t \dot{x}_t \rangle$ which is the term we also obtain.
Another definition is based on the deviation from the fluctuation dissipation relation, the Harada-Sasa relation \cite{Harada2005} used in \cite{Park2020, Chakrabarti2019}, which, similarly to our calculations, gives a linear heat production in the steady state.\\

\modif{The difference of excess heat is given by the difference of variance multiplied by $\kappa/2$, leading to
\be
\Delta Q_{EX} = \frac{D_a}{\omega_0} \frac{\omega_i - \omega_f}{(\omega_0 + \omega_i)(\omega_0 + \omega_f)}
\label{eq:DeltaQEx}
\ee
where we used the notation $\omega_i = \omega_{c, \rm initial}$ and $\omega_f = \omega_{c, \rm final}$.}

\section{Spectral entropy}
\label{APPENDIX_Hs}

\begin{figure}[htb]
	\centering{
		\includegraphics[width=0.6\linewidth]{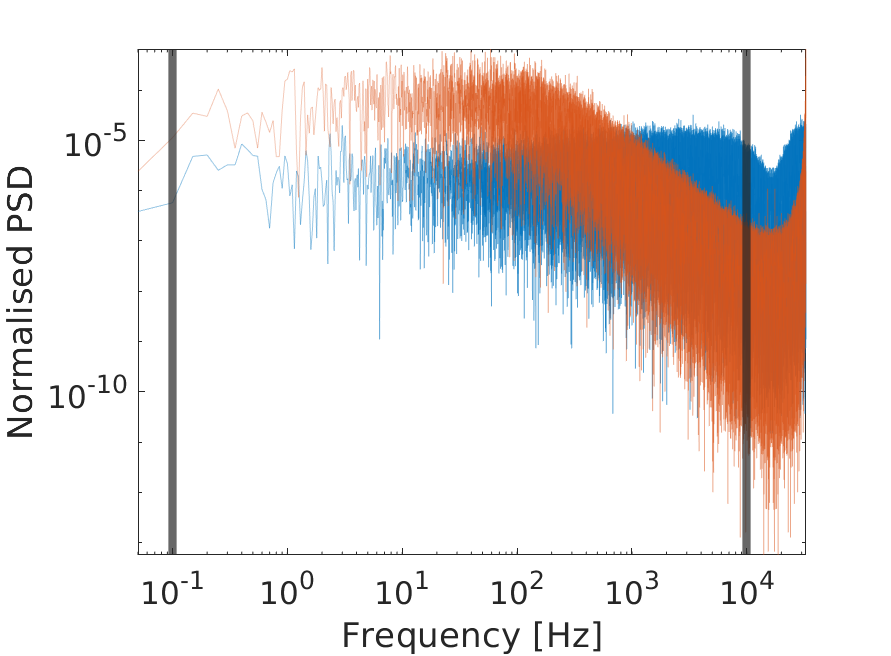}}
	\caption{{Spectra of the white noise (blue line) and colored noise (orange line). The vertical black lines are the limits imposed on the calculation of the spectral entropy at $0.1 \si{\hertz}$ and $10^4\si{\hertz}$.
	}}
	\label{fig:HsCalculation}
\end{figure}

The information content of the injected noise is measured by the spectral entropy $H_S$ \cite{Zaccarelli2013}, which is precisely the Shannon entropy measured in the frequency domain.
To evaluate this quantity and its relation to heat, we perform a series of measurement varying the correlation time of the noise $\tau_c$ while keeping the other parameters constant, and test the robustness of the result for different sets of parameter (stiffness and noise intensity).
For each experiment, an equivalent white noise experiment is also performed, that allows us to extract an effective temperature evaluated through equipartition: $T_{\rm eq} = \kappa \langle x^2 \rangle/k_B$. We then compare $Q_{EX}/k_B T_{\rm eq}$ to $H_S$.
The evaluation of $H_S$ is done on the normalized power spectral density of the noise itself.
In Fig. \ref{fig:HsCalculation}, we represent the normalized power spectral densities (PSD) for  white and colored noises generated at $20~\si{\kilo\hertz}$, along with the spectral boundaries used to get rid of the nonphysical part of the signal (high frequency noise of the electronics).

\vspace{3mm}

On this PSD, the spectral entropy is then evaluated as:
\be
H_S = -\sum_{i=1}^N P(\omega_i) \ln P(\omega_i) ,
\ee
where $P(\omega_i)= S_\eta[\omega_i]/\sum_i  S_\eta[\omega_i]$, and $S_\eta[\omega_i]=2|\eta[\omega_i]|^2$ denotes the PSD of the signal $\eta[\omega]$  at frequency $\omega_i$.

In the main text, we present the data from three different experiments. The first one is set with $\kappa = 33.2~\si{\pico\newton/\micro\meter}$ and a pushing laser maximal power of $150~ \si{\milli\watt}$, leading to a white noise effective temperature of $764.4~\si{\kelvin}$.
The second experiment is performed with $\kappa = 14.8~\si{\pico\newton/\micro\meter}$, pushing laser power $19~\si{\milli\watt}$, leading to $T_{\rm eq} = 531.8~\si{\kelvin}$.
The third experiment is done with $\kappa = 21.4~\si{\pico\newton/\micro\meter}$, pushing laser power $40~\si{\milli\watt}$, leading to $T_{\rm eq} = 943.6~\si{\kelvin}$.
The strong influence of both the stiffness and noise intensity on effective temperature is clear. This influence however does not break the central relation shown in the main text between $Q_{EX}$ and $H_S$.\\

\modif{%
The spectral entropy can be computed analytically in the case studied here of exponentially correlated noise.
We define a normalized pulsation $\Omega = \omega/\omega_c$, and write the normalized spectrum as
\be
P(\Omega) = \frac{S_\eta[\Omega]}{\int S_\eta[\Omega] d\Omega} = \frac{2}{\omega_c} \frac{1}{1 + \Omega^2}.
\ee
The continuous spectral entropy thus writes:
\be
\begin{aligned}
	H_S &= - \omega_c \int P(\Omega) \ln( P(\Omega) ) d\Omega\\
	&= 2 \int \frac{\ln(1+\Omega^2)}{1+\Omega^2} d\Omega + \ln(\omega_c/2) \int \frac{2}{1+\Omega^2} d\Omega\\
	&= 4 \pi \ln 2 + \ln(\omega_c / 2)
\end{aligned}
\ee
where we recognize, for the first integral, the entropy of a Cauchy distribution and, for the second integral, the unit variance of the normalized process.
It leads to a simple expression
\be
H_S = \ln \omega_c + C
\ee
where $C = (4\pi - 1)\ln 2$.
Therefore the difference of spectral entropy between two correlated bath writes as:
\be
\Delta H_S = -\ln(\omega_f/\omega_i).
\label{eq:DeltaHs}
\ee
}\\

\begin{figure}[htb]
	\centering{
		\includegraphics[width=0.6\linewidth]{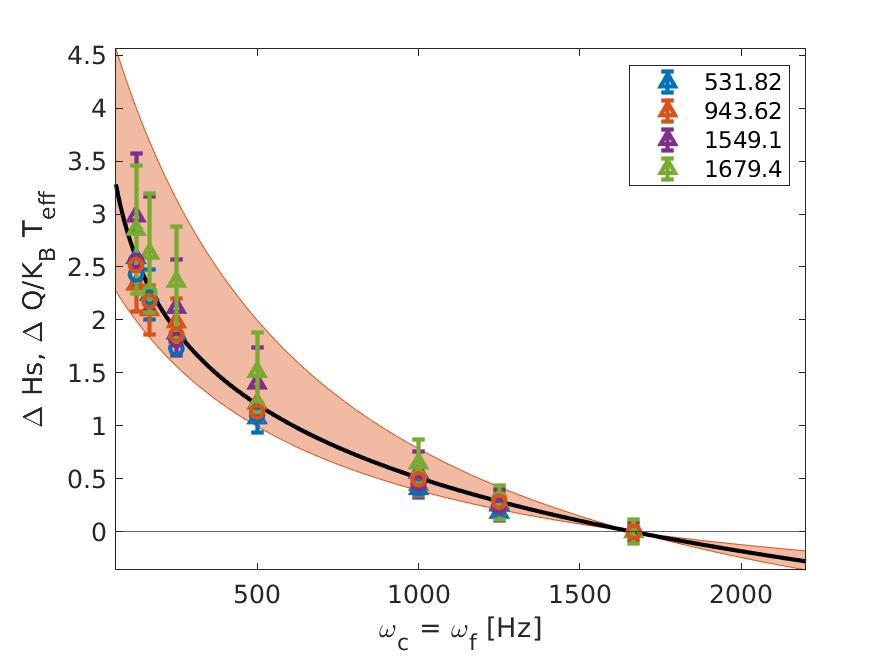}}
	\caption{{\modif{Normalized excess heat difference $\Delta Q_{EX}/ k_B T_{\rm eq}$ (triangles ) and spectral entropy difference $\Delta H_S$ (circles) as a function of the final inverse correlation time $\omega_f$ for various noise amplitudes, each characterized by an equipartition temperature measured in the white noise case. The black solid line corresponds to the analytical expression of the spectral entropy. The red patch corresponds to the ensemble of excess heats, each characterized by a different $D_a$ which is not trivially normalized by $k_B T_{\rm eq}$ as seen Fig. \ref{fig:DaVarVsTauC} (left).}
	}}
	\label{fig:HsDeltaQ}
\end{figure}
\vspace{3mm}

\modif{
On Fig. \ref{fig:HsDeltaQ} we plot the experimental result for excess heat (triangles) and spectral entropy (circles) along with the analytical results Eq. (\ref{eq:DeltaHs}) and Eq. (\ref{eq:DeltaQEx}).
The excess heat explicitly depends on $D_a$ (as seen on Eq. (\ref{eq:DeltaQEx})) which itself depends on the amplitude of the noise (each noise amplitude characterized by an effective temperature $T_{\rm eq}$ will give another $D_a$).
Therefore, the excess heat is not unique, but the various equipartition temperatures $T_{\rm eq}$ give a family of curves, represented here as a shaded red area.
The spectral entropy, in contrast is uniquely defined for all effective temperatures (black solid line).}\\

\modif{We can observe that, despite their very different mathematical expressions (one having a logarithmic profile, while the other is a ratio of polynomials) both functions are very close over a wide frequency bandwidth, making them experimentally indistinguishable within our resolution.
The simplicity of the correlation between heat and information is experimentally true within all our parameter ranges, including noise correlation time and noise amplitude over all probed bandwidths.
}


\bibliography{BiblioColoredNoise}

\end{document}